\documentclass[twocolumn]{aastex631}
\usepackage{amsmath}
\usepackage{cases}

\newcommand\g{\; {\rm g}}
\newcommand\s{\; {\rm s}}

\newcommand\Msun{\; {M}_{\odot}}
\newcommand\kms{\; {\rm km}\,{\rm s}^{-1}}

\newcommand\cm{\;{\rm cm}}

\newcommand\surfunit{\Msun\,{\rm pc^{-2}}}

\newcommand\K{\;{\rm K}}

\newcommand{\W}{\mathcal{W}}
\newcommand{\Peff}{P_\mathrm{eff}}
\newcommand{\tdep}{t_\mathrm{dep}}
\newcommand{\tdyn}{t_\mathrm{dyn}}
\newcommand{\Upstot}{\Upsilon_\mathrm{tot}}
\newcommand{\Ptot}{P_\mathrm{tot}}
\newcommand{\SSFR}{\Sigma_\mathrm{SFR}}
\newcommand{\Sg}{\Sigma_\mathrm{g}}
\newcommand{\Ss}{\Sigma_\mathrm{*}}
\newcommand{\Omegad}{\Omega_\mathrm{d}}
\newcommand{\Omegab}{\Omega_\mathrm{b}}
\newcommand{\rhog}{\rho_\mathrm{g}}
\newcommand{\rhos}{\rho_\mathrm{*}}
\newcommand{\rhob}{\rho_\mathrm{b}}
\newcommand{\rhod}{\rho_\mathrm{d}}
\newcommand{\seff}{\sigma_{\rm eff}}
\newcommand{\Hg}{H_\mathrm{g}}
\newcommand{\Hs}{H_\mathrm{*}}

\newcommand{\PRFMR}{\textit{PRFM-vol}}
\newcommand{\PRFMU}{\textit{PRFM-int}}

\newcommand{\TNG}{\textit{TNG-model}}

\begin{document}

\title{Pressure-regulated feedback-modulated star formation as a subgrid model for galaxy formation simulations}

\author[0000-0002-4232-0200]{Sarah M. R. Jeffreson}
\affiliation{Department of Astrophysical Sciences, Princeton University, 4 Ivy Lane, Princeton, NJ 08544, USA}
\affiliation{Center for Astrophysics, Harvard \& Smithsonian, 60 Garden Street, Cambridge MA, USA}

\author[0000-0002-0509-9113]{Eve C. Ostriker}
\affiliation{Department of Astrophysical Sciences, Princeton University, 4 Ivy Lane, Princeton, NJ 08544, USA}
\affiliation{Institute for Advanced Study, 1 Einstein Drive, Princeton, NJ 08540, USA}

\author[0000-0003-2896-3725]{Chang-Goo Kim}
\affiliation{Department of Astrophysical Sciences, Princeton University, 4 Ivy Lane, Princeton, NJ 08544, USA}

\author[0000-0001-8293-3709]{Jan Burger}
\affiliation{Max-Planck-Institut für Astrophysik, Karl-Schwarzschild-Str. 1, D-85748, Garching, Germany}

\begin{abstract}
We present a new subgrid model for  interstellar gas evolution in cosmological simulations of galaxy formation, based on the pressure-regulated, feedback-modulated (PRFM) theory of star formation. In contrast to the empirically pegged star formation prescriptions employed in current cosmological simulations, the PRFM model links the local star formation rate to the dynamic 
balance achieved in galactic interstellar gas between gravity and stellar feedback effects. With this formulation, both the star formation efficiency and the effective equation of state may be directly calibrated using numerical simulations, such as TIGRESS, which resolve  physics of the interstellar medium and star formation at parsec scales. We develop, and implement in the \textsc{Arepo} moving-mesh code, two complementary classes  of the subgrid model: a volumetric version (\PRFMR) applicable when the gas disk scale height of a galaxy is numerically resolved  in a simulation, and an integrated version (\PRFMU) that reconstructs the mid-plane density and pressure from vertical equilibrium considerations when the true gas scale height cannot be numerically resolved. Using isolated Milky-Way-like disk simulations across mass resolutions $10^5$-$10^7\Msun$, we show that both implementations yield shorter gas depletion times than the IllustrisTNG prescription, especially in regions where pressure and density are large. At high resolution, \PRFMR\ and \PRFMU\ agree closely with each other and with TIGRESS for the star formation rate; \PRFMU\ remains robust at all resolutions tested. These results demonstrate that PRFM-derived subgrid prescriptions provide a physically grounded and numerically stable framework for star formation across the dynamic range of galaxy formation simulations, paving the way for future cosmological applications.
\end{abstract}


\section{Introduction}\label{sec:intro}

Over nearly forty years, hydrodynamic simulations of cosmological volumes, spanning tens to hundreds of megaparsecs, have become important tools for the study of galaxy formation and evolution. Simulation suites based on algorithm frameworks including Illustris 
\citep{Vogelsberger2014},
EAGLE \citep{2015MNRAS.446..521S}, IllustrisTNG \citep{2018MNRAS.475..676S,2018MNRAS.475..648P,2018MNRAS.475..624N},  
SIMBA \citep{2019MNRAS.486.2827D}, and ASTRID \citep{Zhou2025}
(among others) have succeeded in reproducing many of the observed statistical properties of galaxy populations as a function of redshift, providing key insights relating to the roles of stellar and AGN feedback in regulating star formation, the origin of galactic scaling relations, and the thermodynamic and chemical evolution of the circumgalactic medium  \citep[e.g.][]{2015ARA&A..53...51S,Naab2017,2020ARA&A..58..157T,Vogelsberger2020}.

One limitation of large-volume simulations, however, is that it is not computationally feasible to resolve the scales of star-forming giant molecular clouds (GMCs), feedback-driven hot superbubbles, or other substructures  within the multiphase interstellar medium (ISM).  It is therefore not possible to directly model within the ISM the physics, including gas self-gravity, supersonic turbulence, magnetic fields, and several different aspects of stellar feedback, that control the star formation rate (SFR) as well as gas dynamics and thermodynamics. As a result, traditional galaxy formation simulations generally rely on simplified, empirically-calibrated prescriptions for star formation, applied to gas that exceeds a particular threshold in local gas density \citep[e.g.,][]{1992ApJ...399L.113C,Katz92}. Above a threshold, direct heating and cooling is also replaced by an effective equation of state (EoS) which may be motivated by a variety of considerations \citep[e.g.][]{Springel03,2008MNRAS.383.1210S,2013MNRAS.436.3031V}.

Currently, the specific SFR in most large cosmological volumes is assumed to scale with local gas density $\rho$ as $\dot{m}_*/m_\mathrm{g} \propto t_{\rm ff}^{-1} \propto \rho^{1/2}$, where $t_{\rm ff}$ is the free-fall time under gaseous self-gravity.  
The normalization of the adopted star formation prescription is set empirically based on nearby galaxies \citep[e.g.][]{Kennicutt:1998,2012ARA&A..50..531K,2020ARA&A..58..157T}. 
By construction, this yields global gas depletion times $\tdep=M_\mathrm{g}/\dot{M}_*$ that approximately agree with observations at low redshift. However, as discussed in detail in \citet{Hassan2023}, this approach lacks a physical connection to the  processes that actually regulate star formation within the ISM on large scales, and may fail 
in regimes outside that in which
it was calibrated, including  
environments where the density and pressure of gas are very high.  The star formation efficiency is expected to increase in high-pressure environments (see below), and it has been argued 
\citep[e.g.][]{Dekel2023,Mason2023,Andalman2025,Shen2025} that this is key to explaining the abundance of luminous galaxies seen by JWST at high redshift \citep[e.g.][]{Labbe2023,Xiao2024,Casey2024}.

The recent development of high-resolution, full-physics simulations of the star-forming ISM provides a 
new means to calibrate subgrid models for the SFR and  EoS, for application in cosmological simulations of galaxy formation.  This approach has the advantage of being predictive rather than empirically tuned, and allows for environmental parameter regimes that are not present in the local universe.  Optimally, a star formation prescription formulated via this kind of calibration should depend on a limited number of variables that are accessible in a cosmological simulation and robust to varying numerical resolution, while allowing for a large range of galactic conditions.  Developing new subgrid models for star formation that satisfy these requirements, and are supported by both theory and observations, and  is a key objective of the Learning the Universe Collaboration.\footnote{\url{https://learning-the-universe.org/}}

By working within the context of the  pressure-regulated, feedback-modulated (PRFM) theory of star formation \citep[see][and citations within]{Ostriker2022}, it is possible both to identify key physical variables and to derive convenient functional expressions for SFRs that are in agreement with local-universe observations \citep[see][for empirical comparisons]{Ostriker+10,OstrikerShetty2011,Narayanan12, Molina2020,Barrera2021,Sun2020,Sun2023}.  In the PRFM picture, the SFR adjusts so that stellar feedback maintains approximate thermal and vertical dynamical equilibrium in the ISM. Pressure in the ISM responds on the one hand to the weight $\mathcal{W}$ of the ISM in the disk's vertical gravitational field, and on the other hand to 
the energy and momentum injected by supernovae and radiation, proportional to the SFR per unit area $\SSFR$.  
Thus, $\mathcal{W}=\Ptot = \Upstot \, \Sigma_\mathrm{SFR}$ must be simultaneously satisfied for a total ``feedback yield'' $\Upstot$, which leads to a prediction for the SFR in terms of the variables that enter in $\mathcal{W}$.  The dependence of $\Upstot$ on local ISM conditions may be calibrated using high-resolution magnetohydrodynamic simulations, such as those in the TIGRESS and TIGRESS-NCR suites \citep{2017ApJ...846..133K,Kim_CG2020,Ostriker2022, 2023ApJ...946....3K,Kim2024}.  

In recent work, \citet{Hassan2024} applied post-processing to galaxies from the IllustrisTNG simulation suite to investigate how star formation and ISM properties would differ if a PRFM-based prescription, rather than the ``native'' TNG prescription, were adopted. This investigation demonstrated that star-forming galaxies in TNG50 have disks that are vertically well resolved and in dynamical equilibrium,
with low-redshift gas depletion times that are similar to predictions from PRFM.  However, under the high pressure and density conditions that are pervasive at high redshift, the PRFM model predicts shorter gas depletion times than the native TNG model.   \citet{Hassan2024} also compared both the TNG and PRFM models to observations, for the relations $\SSFR$ vs. gas surface density $\Sg$, and $\SSFR$ vs. $\W$. In observations the latter relation, because it takes into account galactic properties beyond just gas content, is tighter than the former \citep{Barrera2021,Sun2023,2024MNRAS.52710201E}. When environment-dependent conversion of CO emission to molecular content is taken into account the prescription $\dot{m}_*/m_\mathrm{g} \propto \rho^{1/2}$ leads to too shallow relations compared to observations \citep[see also][]{Narayanan12,2024ApJ...961...42T}. Motivated by this, in   \citet{Burger2025} an effective EoS based on a TIGRESS calibration was combined with a Schmidt-type star formation prescription  in which the efficiency per free-fall time increases at high gas density ($ \dot{m}_*/m_\mathrm{g} \propto  \rho^1$), for tests of both isolated galaxies and multi-zoom cosmological simulations. Although galaxies in simulations adopting the TIGRESS+Schmidt prescription are somewhat clumpier than counterparts from simulations adopting the prescriptions from TNG or   \citet{Springel03}, low-redshift stellar masses of galaxies are similar. 
The intriguing results from these preliminary investigations motivates a full implementation of the PRFM model + TIGRESS calibration for both star formation and the EoS.

In this paper, we 
extend the proof-of-concept post-processing analysis of \citet{Hassan2024} into a self-consistent, dynamically coupled framework within the moving-mesh code \textsc{Arepo} \citep{2010MNRAS.401..791S}. In our implementation, the functional dependence of the local SFR is derived from the PRFM equilibrium prediction, and both the EoS and star formation yield $\Upstot$ are set from TIGRESS calibrations as a function of the ISM total pressure or weight. 

We have developed, and describe herein, two forms of the PRFM-derived subgrid models for galaxy formation simulations.  One form, \PRFMR, is suitable for simulations in which the vertical thickness of galactic gas disks is resolved, while the other form, \PRFMU, is suitable for coarser resolutions.  \PRFMR\ is similar to traditional star formation prescriptions in terms of being based on local volumetric properties of the galaxy, although not just the gas density but also the density of stars and dark matter enter in defining a local dynamical time.  
Motivated by the requirements for large-box cosmological simulations of galaxies, our goal for \PRFMU\ was to design a star formation subgrid model that is both realistic and robust to changes in  numerical resolution. In the resulting novel approach,
volumetric quantities that are sensitive to numerical resolution are not used directly in the star formation prescription; only integrated quantities that are resolution-insensitive are employed to set $\tdep$ in \PRFMU.

We test both the \PRFMR\ and \PRFMU\  models in isolated, Milky-Way-like disk galaxy simulations across a range of mass resolutions from $10^5$ to $10^7\Msun$.  We compare  SFRs, depletion times, and dynamical times between the two approaches, also comparing to results from simulations 
adopting the IllustrisTNG star formation and EoS prescriptions. Our analysis shows that both the \PRFMR\ and \PRFMU\ implementations at mass resolution $10^5 \Msun$ are able to match the star formation relations obtained in the TIGRESS simulations.  However, the \PRFMR\ implementation begins to show resolution sensitivity in the  $10^6 \Msun$ simulation.  The \PRFMU\ implementation is robust across all resolutions considered here.  

The plan for the remainder of this paper is as follows.  In \autoref{Sec::SF-recipe} we outline the theory behind our new subgrid models (\autoref{sec:PRFM-theory}-\autoref{sec:volumetric}), and describe in detail how the EoS and SFRs are set in the \PRFMR\ (\autoref{Sec::resolved-case}) and \PRFMU\ (\autoref{Sec::unresolved-case}) approaches.  \autoref{sec:mod_results} validates the numerical implementation of the new models (\autoref{sec:valid}), intercompares results for SFRs and timescales from simulations using the \PRFMR, \PRFMU, and \TNG\ implementations  (\autoref{sec:model_comparison}), and delves into detailed properties of the \PRFMR\ (\autoref{Sec::PRFMR-props}) and \PRFMU\ (\autoref{Sec::PRFMU-sims}) simulations at varying resolution to interpret  differences.  Finally, in \autoref{sec:summary} we summarize the main features of the model rollout, as well as the  findings from our initial tests of the models.

\section{PRFM subgrid star formation and ISM model} \label{Sec::SF-recipe}

Here we describe our implementation of 
star formation based on the PRFM theory, with a calibrated feedback yield and effective EoS, as subgrid models for galaxy formation simulations. 

In \autoref{sec:PRFM-theory} we summarize the basic theory, which relies on a relationship between total pressure $\Ptot$ at the midplane and the SFR per unit area $\SSFR$ that is  mediated by stellar feedback. This leads to an expression for the space-and-time averaged gas depletion time $\tdep$ that varies with the characteristic vertical dynamical time $\tdyn$. In turn, 
$\Ptot$
must balance the weight $\W$ of the ISM in equilibrium; this consideration leads to equations for $\tdyn$, along with the gas disk scale height $\Hg$, in terms of local densities (or surface densities) and velocity dispersions. \autoref{sec:parameters} summarizes how stellar and dark matter parameters that enter in equilibrium expressions for $\Hg$ and $\tdyn$ relate to locally measurable galaxy properties.  

In \autoref{sec:TIGRESS-calib}, we provide expressions for key quantities in the PRFM theory, the total feedback yield $\Upstot$ and the effective velocity dispersion $\seff$ based on the EoS relating total effective pressure to density, which are calibrated based on TIGRESS simulations. \autoref{sec:volumetric} explains how vertically-integrated PRFM relations are translated to a volumetric star formation relation that can be implemented as a subgrid model in a galaxy formation simulation in which the gas disk is vertically resolved. 

In \autoref{Sec::resolved-case}, we describe our ``volumetric'' (\PRFMR) 
implementation of the subgrid model in the \textit{Arepo} code. \PRFMR\ is appropriate at higher resolutions, equal to or finer than the resolution of the TNG50 simulation, in which the scale-heights of both the gas and stellar components of disk galaxies are resolved \citep{2019MNRAS.490.3196P}.
\autoref{Sec::unresolved-case}  then describes our ``integrated'' (\PRFMU) implementation in \textit{Arepo}, appropriate at lower resolutions comparable to the TNG100 and TNG300 simulations (see {\tt https://www.tng-project.org/about/}), in which the scale-heights of both the gas and stellar components of disk galaxies remain unresolved.

\subsection{PRFM Theory}\label{sec:PRFM-theory}

Star formation can be characterized on many different spatial and temporal scales.  When averaging over a sufficiently large spatial and/or long temporal scale, the full lifecycle of star-forming, multiphase interstellar gas is well represented, and a quasi-equilibrium rate is applicable.   
For any given (vertically-integrated) patch of the galaxy's gas disk, the SFR can then be defined using a gas depletion time $\tdep$ or equivalently an efficiency per dynamical time $\varepsilon_\mathrm{dyn}$ as
\begin{equation}\label{eq:SFR}
    \frac{\dot M_*}{M_\mathrm{g}} =\frac{\SSFR}{\Sg} \equiv \frac{1}{\tdep} \equiv \frac{\varepsilon_\mathrm{dyn}} {\tdyn}; 
\end{equation}
Here, $\dot M_* = A \SSFR$ and $M_\mathrm{g}=A\Sg$ for $\SSFR$ and $\Sg$ the SFR and gas mass per unit area and $A$ the area of a patch that is large enough to fully sample from all lifecycle stages.  The dynamical time above is defined as  
\begin{equation} \label{Eqn::tdyn_general}
    \tdyn\equiv \frac{2 \Hg}{\seff} 
\end{equation}
for $\seff\equiv (\Ptot/\rhog)^{1/2}$ the effective velocity dispersion and $\Hg$ the half-thickness of the gas disk, which is defined as $\Hg\equiv \Sg/(2\rhog)$ for  $\rhog$ the midplane gas volume density; $\Ptot$ is the midplane pressure.  
We then have 
\begin{equation}\label{eq:pressure}
 \Ptot = \sigma_\mathrm{eff}^2 \rhog  =  \sigma_\mathrm{eff}^2 \frac{\Sg}{2\Hg} = \seff \frac{\Sg}{\tdyn}.
\end{equation}
The formulation in \autoref{eq:SFR} of star formation in terms of an efficiency 
per dynamical time is analogous  to formulations in terms 
of an efficiency per free-fall time (with $t_\mathrm{ff}=[3\pi/(32G\rhog)]^{1/2}$), but 
$\tdyn$ is a more general dynamical time, which does 
not assume that gas gravity is dominant in confining
the ISM on large scales and setting $\Hg$.  

In the PRFM theory \citep[see][and references therein]{Ostriker2022}, under the assumption of quasi-equilibrium, the total ISM pressure at the midplane of a disk galaxy  is related to the SFR per unit area by 
\begin{equation}\label{eq:P_SFR}
\Ptot= \Upstot\SSFR,
\end{equation}
where the feedback yield $\Upsilon_\mathrm{tot}$
is determined by balancing various forms of energy gains with dissipation and losses.  The pressure in general includes thermal, turbulent, and magnetic terms,  so that $\Upsilon_\mathrm{tot}$ is the sum of these individual feedback yield terms.  
Because $\Upstot$ depends primarily on stellar and atomic physics and photoprocesses, it is a weak function of galaxy parameters so that $\Ptot$ varies nearly linearly with $\SSFR$ \citep{Ostriker+10,OstrikerShetty2011,Kim13,KimCG&Ostriker15a}; we shall refer to  $\Ptot$ vs. $\SSFR$ as an ``Ostriker-Kim'' relation, with numerical calibrations for $\Upstot$ from TIGRESS simulations discussed in \autoref{sec:TIGRESS-calib} and \autoref{app:A}.

From \autoref{eq:SFR}, \autoref{eq:pressure}, and \autoref{eq:P_SFR}, it is straightforward to show that 
\begin{equation} \label{Eqn::tdep_general}
\tdep = \frac{\Upstot \Sg}{\Ptot}= \frac{2\Upstot \Hg}{\seff^2}=\frac{\Upstot}{\seff}\tdyn, 
\end{equation}
which is equivalent to a star formation efficiency per dynamical time of
\begin{equation}
\varepsilon_\mathrm{dyn} \equiv \frac{\tdyn}{\tdep}= \frac{\sigma_{\rm eff}}{\Upsilon_\mathrm{tot}}.     
\end{equation}
In general, $\seff/\Upstot\ll 1$, so the star formation efficiency is modest, although it increases under high density (and pressure) conditions.  Although in many cosmological subgrid models 
the analogous efficiency per free-fall time is set to a constant that is
independent of the local conditions, 
in the PRFM theory $\seff$ and $\Upstot$ vary with 
ISM and galactic conditions, and therefore 
$\varepsilon_\mathrm{dyn}$
is not constant (see \autoref{eq:epsdyn_calib} below).

For a disk galaxy under the assumption of vertical dynamical equilibrium, the total midplane pressure $\Ptot$
will be equal to the weight $\cal W$ of the ISM in the total gravitational potential, which in general includes terms from the gas, the stellar disk, the stellar bulge, and the dark matter halo.\footnote{Strictly speaking, it is the difference $\Delta \Ptot$ in total pressure across the disk that is balanced by the ISM weight $\mathcal{W}$.  While cosmic rays are similar to other pressures in having $P_\mathrm{cr}\propto \SSFR$, their scale height is very large and they therefore do not contribute to balancing $\mathcal{W}$. Thus, when using $\mathcal{W}=\Ptot = \Upstot \SSFR$ to compute $\tdep= \Upstot\Sg/\mathcal{\Ptot}=2 \Upstot \Hg/\seff^2$, the cosmic ray contribution is not included in $\Upstot$ or in $\seff$.}
One can show \citep{Ostriker+10,OstrikerShetty2011,Hassan2024} that 
the ISM weight based on the sum of these terms is given by 
\begin{equation}\label{eq:weight_total}
\mathcal{W} = \frac{\pi G \Sg^2}{2} \left[ 1 + \frac{\Sigma_*}{\Sg} \frac{2\Hg}{\Hg+H_*} + \frac{2 \zeta (\Omegad^2 +\Omegab^2)}{\pi G \Sg} \Hg\right],    
\end{equation}
where
the gas disk's equilibrium half thickness $\Hg$ can be obtained by setting  \autoref{eq:weight_total} equal to the third term in \autoref{eq:pressure} and solving  
\begin{equation}\label{eq:combined_Hg_eq}
  \Hg\left[ 1 + \frac{\Sigma_*}{\Sg} \frac{2\Hg}{\Hg+H_*} + \frac{2 \zeta (\Omegad^2 +\Omegab^2)}{\pi G \Sg} \Hg\right] = \frac{\seff^2}{\pi G \Sg}. 
\end{equation}
In these expressions, $\Ss$ and $\Hs$ are the surface density and half-thickness of the stellar disk, $\Omegad^2 = r^{-1} d\Phi_d/dr$ and $\Omegab^2 = r^{-1} d\Phi_b/dr$ are the contributions to angular velocity from spherical dark matter and bulge potentials, and $\zeta \approx 1/3$.  \autoref{eq:combined_Hg_eq} leads to a cubic equation for $\Hg$, 
\begin{eqnarray}\label{eq:Hgas_cubic_Sigma}
    \Hg^3 \left(  \frac{2\zeta[\Omega_\mathrm{d}^2 +\Omegab^2] }{\pi G \Sigma_g} \right) +
    \Hg^2 \left(1 + \frac{2\Sigma_*}{\Sg} + \frac{2\zeta[\Omega_\mathrm{d}^2 +\Omegab^2] H_*}{\pi G \Sg} \right)\nonumber\\  
    + \Hg \left(H_* - \frac{\sigma_{\rm eff}^2 }{\pi G\Sg} \right) - H_* \frac{\sigma_{\rm eff}^2 }{\pi G\Sg}=0,\hskip 1cm 
\end{eqnarray}
with solution given in Equation (20) of \citet{Hassan2023}.\footnote{The expressions for coefficients in Equations 16-19 of \citet{Hassan2023} omit the bulge for simplicity; to allow for a bulge, the simple substitution $\Omegad^2\rightarrow \Omegad^2 + \Omegab^2$ is made.}  For either an observation or simulation in which $\Sg$, $\Ss$, 
$\Omegad$, 
$\Omegab$, $\seff$, and $\Hs$ can be directly measured or estimated, the solution of \autoref{eq:Hgas_cubic_Sigma} provides a prediction for the equilibrium $\Hg$.

In some circumstances it may be appropriate to treat the ratio $\Hs/\Hg$ in \autoref{eq:combined_Hg_eq} as a parameter, which
leads to a quadratic that may be solved for $\Hg$, yielding:  
\begin{eqnarray}\label{eq:H_Sigma}
\Hg = \frac{2\seff^2}{\pi G[\Sg + \Sigma_*2/(1+\Hs/\Hg)]} \times \hskip1.2in \nonumber \\ 
\left(1 
+ \left[
1 +
 \frac{8\zeta (\Omega_\mathrm{d}^2 + \Omegab^2)\seff^2}{(\pi G)^2[\Sg + \Sigma_*2/(1+\Hs/\Hg)]^2}\right]^{1/2}\right)^{-1}.
 \hskip0.6cm
\end{eqnarray}
An alternative form can be 
obtained by substituting  $\Sg\rightarrow 2\Hg \rhog$ and  $\Ss\rightarrow 2\Hs \rhos$ in
\autoref{eq:combined_Hg_eq},
leading to 
\begin{equation}\label{eq:H_rho}
\Hg= \frac{\seff}{\left[2\pi G \rhog + \frac{4 \pi G \rho_*}{1+\Hg/H_*} + 2 \zeta (\Omega_\mathrm{d}^2 + \Omegab^2)\right]^{1/2}}.   
\end{equation}
Given direct measurements of either ($\Sg,\Ss$) or  ($\rhog,\rhos$), $\Omegad$, $\Omegab$, and a measurement or calibration of $\Hg/\Hs$, the value of $\Hg$ may be obtained from either \autoref{eq:H_Sigma} or \autoref{eq:H_rho}. 

Once $\Hg$ has been obtained from  \autoref{eq:Hgas_cubic_Sigma}, \autoref{eq:H_Sigma}, or \autoref{eq:H_rho}, the (patch-averaged) dynamical time 
is obtained as $\tdyn=2\Hg/\seff$  (from \autoref{Eqn::tdyn_general}), 
 and the equilibrium midplane density and pressure are given as $\rhog=\Sg/(2\Hg)$ and $\Ptot=\seff^2\Sg/(2\Hg)$ (from \autoref{eq:pressure}). Provided a presciption for $\Upstot$ (which has been calibrated as a function of $\Ptot$; see \autoref{sec:TIGRESS-calib}), the (patch-averaged) depletion time is then $\tdep= (\Upstot/\seff)\tdyn$ (from 
\autoref{Eqn::tdep_general}). 

\subsection{Galactic Component Parameter Values}\label{sec:parameters}

The bulge contribution to the vertical gravity and hence the equilibrium $\Hg$ and $\tdyn$ depend on the 
stellar profile within the bulge, generally 
taking the form
\begin{equation}
\Omegab^2= 2 \pi G  a_b   \rhob 
\end{equation}
where $a_b$ is an order-unity dimensionless coefficient and $\rhob$ is the stellar density of the bulge.
For a constant-density bulge, $a_b = 2/3$, and 
for a Hernquist bulge, $a_b$ varies from 1 to 2.
Thus, for a Hernquist bulge, the mean value of $2\zeta \Omegab^2 \rightarrow 2 \pi G \rhob$. We shall adopt 
this form in the present work, but the 
numerical factor $a_b$ may be calibrated for the general case.

The contribution from the dark matter halo to the vertical gravity will also depend on its density profile.  For a flat rotation curve, $\Omegad^2 \rightarrow 
4 \pi G \rhod$, but the NFW profile is more realistic in the star-forming regions of galaxies. For the NFW profile, $\Omegad^2 \rightarrow 2 \pi G \rhod $ inside the scale radius $R_s$, so that $2 \zeta  \Omegad^2 \rightarrow (4\pi /3) G \rhod$.  

Within much of the star-forming disks of galaxies in the nearby universe, realistic values of $\Hg/\Hs$  span a relatively small range $\sim 0.3-1$, although $\Hg/\Hs$ drops approaching the bulge-dominated region \citep[e.g.][]{Vijayakumar2025}. 
At high redshift, where stellar 
disk thickening has not yet significantly developed, $\Hg/H_*\approx1$ is a reasonable zeroth order approximation.  More generally, from mutual vertical equilibrium of both a gaseous and stellar disk, we may expect $H_\mathrm{g}/H_\mathrm{*}\sim\sigma_\mathrm{eff}/\sigma_\mathrm{*,z}$ for $\sigma_\mathrm{*,z}$ the vertical velocity dispersion of the stellar component \citep[e.g.][]{Elmegreen1989}. \autoref{eq:H_rho} for $\Hg$ is insensitive to the exact the value of $\Hg/\Hs$ provided this ratio is $\lesssim 1$.  When  \autoref{eq:H_Sigma} is used,  $\Hg$ and $\tdyn$ could be underestimated  
if $\Ss/\Sg \gg 1$ and an unrealistically small value of $\Hs/\Hg$ is used. 

Taken together, the above considerations imply using \autoref{eq:H_rho} in $\tdyn =2\Hg/\seff$ 
that 
\begin{equation}\label{eq:tdynest}
\begin{split}
\tdyn&\approx \frac{2}{(2\pi G)^{1/2}}\left[\rhog +\rhos \frac{2}{1+ \Hg/H_*} + \rhob +\frac{2}{3} \rhod\right]^{-1/2},\\
\end{split}   
\end{equation}
where  $\seff/\sigma_{*,z}$ may be used instead of $ \Hg/H_*$  if the scale height ratio cannot be measured or estimated accurately.  
\autoref{eq:tdynest} shows that all mass components contribute 
to $\tdyn$ in a similar way, such that it will be primarily set by the component
with the highest local density. In the limit when gas dominates the density, $\tdyn = 1.47 t_\mathrm{ff}$.

\subsection{Feedback Yield and EoS functions}\label{sec:TIGRESS-calib} 

From the TIGRESS numerical simulations, we have calibrations for $\Upstot$ as a function of the total ISM pressure $P_\mathrm{tot}$, and also a calibration of the effective EoS (i.e.~$\Ptot$ as a function of density $n_H$).  The latter is equivalent to a functional dependence of $\seff$ on $\Ptot$. 

From \citet{Ostriker2022}, the calibration for the feedback yield $\Upstot \equiv \Ptot/\SSFR$ is given by 
\begin{equation} \label{Eqn::upsilon}
\Upstot = 1.03 \times 10^3 \kms\left(\frac{P_{\rm tot}}{10^4 k_{\rm B} {\rm cm}^{-3} {\rm K}}\right)^{-0.21},
\end{equation}
and the calibration for the effective EoS is given by
\begin{equation} \label{Eqn::prfm-eos}
P_{\rm tot}= 4.7 \times 10^4 k_{\rm B} {\rm cm}^{-3} {\rm K}\left(\frac{n_{\rm H}}{{\rm cm}^{-3}}\right)^{1.8},
\end{equation}
such that
\begin{equation} \label{Eqn::sigma_eff}
\sigma_{\rm eff}= 12 \kms\left(\frac{P_{\rm tot}}{10^4 k_{\rm B} {\rm cm}^{-3} {\rm K}}\right)^{0.22}.
\end{equation}
Combining \autoref{Eqn::upsilon} with \autoref{Eqn::sigma_eff}, we obtain a calibrated function for the (patch-averaged) star formation efficiency per dynamical time,
\begin{equation}\label{eq:epsdyn_calib}
    \varepsilon_\mathrm{dyn} =\frac{\seff}{\Upstot} = 0.012 \left(\frac{P_{\rm tot}}{10^4 k_{\rm B} {\rm cm}^{-3} {\rm K}}\right)^{0.43}.
\end{equation}

Because pressure increases in the high density regions of galaxies and at high redshift overall, 
\autoref{eq:epsdyn_calib} predicts that 
star formation is expected to be more efficient under these conditions. 
From \autoref{Eqn::tdep_general} and \autoref{eq:tdynest}, the above calibrations from TIGRESS imply the  (vertically averaged) depletion time would approximately follow 
\begin{equation}
\label{eq:tdep_explicit}
    t_{\rm dep} 
    \sim 0.5\,{\rm Gyr }
    \left(\frac{n_\mathrm{H}}{{\rm cm^{-3}}}\right) ^{-0.77}\left(\frac{\rho_\mathrm{baryon}}{\Msun\,{\rm pc^{-3}}} +  \frac{(2/3)\rhod}{\Msun\,{\rm pc^{-3}}}\right)^{-1/2} 
\end{equation}
where $\rho_\mathrm{baryon}=\rhog + \rho_* +\rhob$ is the sum of the gas, stellar disk, and stellar bulge densities at the disk midplane, and we have assumed $\seff/\sigma_{*,z}\sim 1$. \autoref{eq:tdep_explicit} has a considerably steeper dependence on density than would be the case if the efficiency per free-fall time in the gas were taken to be constant, which leads to $\tdep \propto n_\mathrm{H}^{-1/2}$.

The calibrations given in \autoref{Eqn::upsilon}, 
\autoref{Eqn::prfm-eos}, and 
\autoref{Eqn::sigma_eff}
above are based on simulations with varying dynamical conditions but fixed (solar neighborhood) metallicity. \citet{Kim2024} considered a similar range of dynamical conditions and also allowed for varying metallicity, employing an updated version of the TIGRESS framework with direct ray-tracing radiation.  The fits obtained for the feedback yield and effective velocity dispersion from these simulations are summarized in \autoref{app:A}.

\subsection{Volumetric PRFM Star Formation Model}
\label{sec:volumetric}

The PRFM theory outlined above provides predictions for the gas depletion time within a patch of the ISM disk.  However, for the purposes of implementation as a star formation subgrid model, it is necessary to provide an expression for the depletion time in every fluid element of a simulation, rather than a depletion time that vertically integrates and horizontally averages over fluid elements.  

The PRFM theory does not in itself constrain the functional form for the volumetric SFR law; rather it imposes a constraint on the vertical average of the volumetric law.
Namely, if $\dot{\rho}_*$ is the local volumetric SFR, from \autoref{eq:SFR} and \autoref{Eqn::tdep_general} we must have 
\begin{equation}\label{eq:SFRconstraint}
  \frac{\int dz\, \rhog }{\int dz\, \dot{\rho}_*} =  \frac{\Upstot}{\seff}\tdyn,
\end{equation}
where $\Upstot$ and $\seff$ are evaluated via the calibrations in \autoref{Eqn::upsilon} and \autoref{Eqn::sigma_eff} using the midplane pressure for $\Ptot$, and $\rhog$ and $\dot{\rho}_*$ are understood as horizontally-averaged over a large enough scale to capture local ISM and star formation variations.  The ``patch-averaged'' dynamical time $\tdyn$ is defined in \autoref{Eqn::tdyn_general}, and may be evaluated using midplane densities via \autoref{eq:tdynest}. 

Given that the functional form of $\dot{\rho}_*$ is not constrained, the simplest possible approach is to assume that the \textit{local} depletion time has the same functional dependence on local gas pressure (through the ratio $\Upstot/\seff$) and local stellar, gas, and dark matter densities  (through a ``local'' version of \autoref{eq:tdynest}) as the ``patch-averaged''  depletion time in \autoref{Eqn::tdep_general}, but with an adjusted normalization.  
Under this assumption, we have 
\begin{equation}\label{eq:SFRvol}
\begin{split}
    \frac{\dot{\rho}_*}{\rhog} \equiv \frac{1}{\tdep}
    &= \, \mathcal{R}_f  \frac{\seff}{\Upstot} \frac{1}{t_\mathrm{dyn,local}} \\
    &=\mathcal{R}_f  \frac{\sigma_0}{\Upsilon_0}  \left( \frac{P_\mathrm{eff}}{P_0}\right)^{\alpha + \beta} \frac{1}{t_\mathrm{dyn,local}}\\
\end{split}
\end{equation}
where $\seff$ and $\Upstot$ are evaluated at the local effective pressure $P_\mathrm{eff}$ in a given cell of the cosmological simulation (set via a subgrid EoS), and $\mathcal{R}_f$ is a renormalization factor (to be determined).  In the second line, we have assumed functional forms 
$\Upstot=\Upsilon_0 (P_\mathrm{eff}/P_0)^{-\alpha}$ and $\seff =\sigma_0(P_\mathrm{eff}/P_0)^\beta$ based on calibrations 
from resolved ISM simulations as in \autoref{Eqn::upsilon} and \autoref{Eqn::sigma_eff} (for 
$P_0 = 10^4 k_{\rm B} {\rm cm}^{-3} {\rm K}$ the pressure unit adopted in these expressions).   
Employing the functional form of  \autoref{eq:tdynest} to evaluate  $t_\mathrm{dyn,local}$ using densities local to a given cell, and defining $\varepsilon_0=\mathcal{R}_f \sigma_0/\Upsilon_0$, we can write
\begin{equation}\label{eq:SFRvol_den}
\begin{split}
 \frac{\dot{\rho}_*}{\rhog} =\, \varepsilon_0 &\left( \frac{P_\mathrm{eff}}{P_0}\right)^{\alpha + \beta} \left(\frac{\pi G}{2} \right)^{1/2} \\
 \times  &\left[ \rhog + \frac{2}{1+ \Hg/H_*}  \rho_*   + \frac{2}{3}\rhod \right]^{1/2}.
\end{split}
\end{equation}
The  renormalization factor $\mathcal{R}_f$ in the coefficient can then be determined from the constraint in \autoref{eq:SFRconstraint}.  Note that we have not explicitly separated out the bulge and disk components of the stellar density in writing the second line of \autoref{eq:SFRvol_den}, for reasons discussed in \autoref{sec:parameters}.  

It is often the case that the stellar and gas density have comparable vertical scale heights and the dark matter density is sub-dominant.  In this case, \autoref{eq:SFRvol_den} becomes 
\begin{equation}\label{eq:rhostardot}
  \dot{\rho}_*  =  \varepsilon_0 \left( \frac{P_\mathrm{mid}}{P_0}\right)^{\alpha + \beta} \frac{\rho_\mathrm{g,mid}}{t_\mathrm{dyn,mid}} s^{\frac{\alpha +\beta}{1-2\beta} + \frac{3}{2}}
\end{equation}
where $s=\rhog/\rho_\mathrm{g,mid}$ is the gas density relative to the midplane value and we have used $P_\mathrm{eff}/P_\mathrm{mid} = (\rhog/\rho_\mathrm{g,mid})^{1/(1-2\beta)}$ from the EoS. Then applying \autoref{eq:rhostardot} in \autoref{eq:SFRconstraint} leads to 
\begin{equation}
    \varepsilon_0= \frac{\sigma_0}{\Upsilon_0} \frac{\int dz s}{\int dz s^{\frac{\alpha +\beta}{1-2\beta} + \frac{3}{2}} }.
\end{equation}
In the case of an exponential vertical density profile with $s_\mathrm{th}$ the density ratio at the threshold for star formation, evaluating the integrals leads to
\begin{equation}\label{vol_coeff}
  \varepsilon_0= \frac{\sigma_0}{\Upsilon_0}\left(\frac{\alpha +\beta}{1-2\beta} + \frac{3}{2}  \right) \frac{1-s_\mathrm{th}}{1-s_\mathrm{th}^{ \frac{\alpha +\beta}{1-2\beta} + \frac{3}{2} } }.    
\end{equation}
In a marginally resolved simulation, $s_\mathrm{th}= 1 -\Delta$ for $\Delta$ small, so that $\varepsilon_0 = \sigma_0/\Upsilon_0$, i.e. $\mathcal{R}_f=1$.  Thus, when resolution is marginal, we recover the ``patch-averaged'' relationship.  
In a well resolved simulation, however, $s_\mathrm{th}\ll 1$ so that 
\begin{equation}\label{eq:renorm_fac}
  \varepsilon_0= \frac{\sigma_0}{\Upsilon_0}\left(\frac{\alpha +\beta}{1-2\beta} + \frac{3}{2}  \right)  .    
\end{equation}
Adopting $\alpha= 0.21$ from the index in \autoref{Eqn::upsilon} and $\beta = 0.22$ from the index in \autoref{Eqn::sigma_eff}, we can see that the coefficient in the volumetric star formation relation would need to be increased by a factor $\mathcal{R}_f=2.3$ compared to the coefficient in the vertically integrated, patch-averaged relation (cf.~\autoref{Eqn::tdep_general}).  For a Gaussian vertical density profile, if the disk is well resolved the renormalization factor would instead be $\mathcal{R}_f= \left[(\alpha + \beta)/(1-2\beta) + 3/2  \right]^{1/2}$, equal to 1.5 for our values of $\alpha$ and $\beta$.  We shall adopt $\varepsilon_0=2\sigma_0/\Upsilon_0$, i.e. $\mathcal{R}_f=2$, for our simulations with the \PRFMR\ implementation. 

It is worth noting that with \autoref{eq:SFRvol_den}, the volumetric SFR, and therefore the density distribution of newly formed stars, declines more steeply than the gas density with offset from the midplane $|z|$.  For example, with \autoref{eq:SFRvol_den} and $\alpha=0.21$, $\beta=0.22$, the scale height of star formation $H_\mathrm{SF}$ would be smaller than the gas scale height $\Hg$ by a factor $\mathcal{R}_f=2.3$  or $1.5$ for an exponential or Gaussian density profile, respectively.  Having $H_\mathrm{SF}/\Hg<1$ is consistent with expectations, when $\Hg$ is taken to be equivalent to the gas scale height averaged over all phases. However,  both observations and numerical simulations with high numerical resolution and the realistic microphysics needed to capture substructure of the cold ISM typically find a greater contrast, $\Hg/H_\mathrm{SF}\sim 3-4$ \citep[e.g.][]{1994ApJ...433..687M,1995ApJ...448..138M,2020ApJ...898...35K,2021ApJ...910..131S,2023ARA&A..61...19M,2022MNRAS.515.1663J,2024ApJ...975..173L}.  

In principle, it would be possible to impose an explicit vertical distribution of the SFR, while requiring \autoref{eq:SFRconstraint} is satisfied.  For example, we could impose $\dot\rho_*\propto e^{-z/H_\mathrm{SF}}$ with $H_\mathrm{SF}/\Hg$ calibrated from simulations,  which for an exponential density profile would lead to expressions analogous to  \autoref{eq:rhostardot} and \autoref{eq:renorm_fac} with $(\alpha+\beta )/(1-2\beta) + 3/2\rightarrow \Hg/H_\mathrm{SF}$.  However, this would require an additional measurement of $|z|$ for each volume element, which would be costly in a cosmological simulation. More generally, in the future it will be of interest to explore  alternative volumetric prescriptions to that in \autoref{eq:SFRvol} that satisfy \autoref{eq:SFRconstraint} and also allow for parameterization of $H_\mathrm{SF}/\Hg$.  This would presumably require more than just local $\rhog$ and $P_\mathrm{eff}$ as inputs, since it would represent an additional degree of freedom.

\subsection{Subgrid Model When Disk Scale-Heights Are Resolved (\PRFMR)} 
\label{Sec::resolved-case}

In this section, we describe the implementation of a subgrid model for the EoS and SFR in Arepo, based on the prescription outlined in \autoref{sec:PRFM-theory},  \autoref{sec:TIGRESS-calib}, and \autoref{sec:volumetric}, in the ``resolved'' case for which both the gas and stellar disk scale-heights are well-resolved in the vertical direction.

First, for each gas cell in the simulation that exceeds a star formation density threshold of $n_\mathrm{th}=\rho_\mathrm{g,th}/(1.4 m_H) = 0.13\, {\rm cm}^{-3}$, we set the pressure in each cell to an effective value $P_\mathrm{eff}$ based on the TIGRESS calibration. Because the disk is expected to be vertically resolved, we may use the measured volume density $\rhog$ in each cell, at simulation run-time, to set $n_{\rm H} = \rhog/(1.4 m_{\rm H})$, and then set $P_\mathrm{eff}$ equal to $\Ptot$ as predicted at the cell's $n_{\rm H}$ using \autoref{Eqn::prfm-eos}, with the cell's specific energy equal to $P_\mathrm{eff}/[ (5/3) \rhog]$. The density threshold $n_\mathrm{th}$ is chosen to coincide with the Illustris TNG threshold for star formation~\citep{2018MNRAS.475..624N}, for later purposes of comparison.

This EoS is applied in lieu of updating the energy via direct integration of the energy equation with source terms, because the true heating (by feedback from star formation) and cooling (requiring detailed ISM photochemistry) cannot be directly implemented at the resolution of cosmological simulations.  The effective pressure and specific energy also reflect the subgrid model for turbulence and magnetic fields driven by feedback.

In order to set the SFR of individual Voronoi gas cells, we need to compute the depletion time. Following the PRFM formulation, and allowing for 
a renormalization factor $\mathcal{R}_f=2$ to increase the SFR in each Voronoi cell relative to the values from the ``patch-averaged'' calibration (see \autoref{sec:volumetric}), we apply 
\autoref{eq:SFRvol}.  The feedback yield $\Upstot$ and effective velocity dispersion $\seff$ are calculated in terms of the local effective pressure $P_\mathrm{eff}$ (which itself is set from the calibrated EoS function) via the TIGRESS calibrations in \autoref{Eqn::upsilon} and \autoref{Eqn::sigma_eff}. 

To maintain consistency between $P_\mathrm{eff}$, $\sigma_{\rm eff}$, and $\Upstot$ at all simulation times, we set \autoref{Eqn::upsilon}--\autoref{Eqn::sigma_eff} simultaneously. In the case of a gas cell that exceeds the density threshold for star formation, but for which $P_\mathrm{eff} < 10^4 k_{\rm B} {\rm cm}^{-3} {\rm K}$, we set $\sigma_{\rm eff} = 12~{\rm km/s}$, which is continuous with \autoref{Eqn::sigma_eff}.


To evaluate $t_\mathrm{dyn}$, in the case of a vertically resolved disk for which the measured volume densities of gas, stars, and dark matter are resolved, \autoref{eq:H_rho} may be used for $\Hg$ in 
$t_\mathrm{dyn}=2 \Hg/\seff$ (here and henceforth, for cleaner notation we omit the ``local'' subscript in the dynamical time). In turn, \autoref{eq:H_rho} requires values for the stellar and dark matter volume densities, $\rhos$ and $\rhod$, which we compute simultaneously by walking the gravity/force tree within a fixed search radius of $500$~pc around each star-forming gas cell. We weight the masses of the stellar and dark matter particles within each search radius using a cubic spline kernel centered on the gas cell centroid.

We note that is not so straightforward to separate the ``bulge'' and ``disk'' portions of the total stellar density $\rho_\mathrm{stellar}$ within the simulation volume. One method that has been adopted is to compute the density of all stars that are counter-rotating and assign twice this value (up to $\rho_\mathrm{stellar}$) to the bulge as $\rhob$, while assigning the remainder to the disk as $\rho_*=\rho_\mathrm{stellar}-\rhob$~\citep{2015ApJ...804L..40G}. With this or a related approach, it is possible to separately include the bulge and disk contributions in \autoref{eq:H_rho}.  Since, however,  the numerical coefficients are in fact quite similar (see \autoref{sec:parameters}), and our test focuses on a disk-dominated galaxy, here we shall dispense with this small refinement. 

Since we do not explicitly separate out the bulge term, 
from \autoref{eq:tdynest} 
for the dynamical time in the vertically resolved case we have 
\begin{equation}\label{eq:tdynrho}
\tdyn=  \frac{2}{\left[2\pi G \rhog + \frac{4 \pi G \rho_*}{1+\Hg/H_*} + \frac{4\pi}{3}  G\rhod\right]^{1/2}}.  
\end{equation}
If we wish to use \autoref{eq:tdynrho}, we also require a value for $\Hg/H_*$.
Adopting  $\zeta  \Omegad^2 \rightarrow (2\pi /3) G \rhod$ and dropping the explicit bulge term, \autoref{eq:Hgas_cubic_Sigma} for $\Hg$  becomes 
\begin{equation} \label{Eqn::Hg_cubic_resolved_Sigma}
\begin{split}
H_{\rm g}^3 \Big(&\frac{4 \rhod}{3\Sigma_{\rm g}} \Big) + H_{\rm g}^2 \Big(1 + \frac{2\Sigma_*}{\Sigma_{\rm g}} + \frac{4\rho_{\rm d} H_*}{ 3\Sigma_{\rm g}}\Big) \\
&+ H_{\rm g}\Big(H_* - \frac{\sigma_{\rm eff}^2}{\pi G \Sigma_{\rm g}}\Big) - H_* \frac{\sigma_{\rm eff}}{\pi G \Sigma_{\rm g}} = 0.
\end{split}
\end{equation}
Once we have solved the cubic equation (\autoref{Eqn::Hg_cubic_resolved_Sigma}) to compute $H_{\rm g}$, we substitute its value into \autoref{eq:tdynrho} to compute $t_{\rm dyn}$ and thus $t_{\rm dep}=\tdyn\Upstot/(\mathcal{R}_f \seff)$, setting the cell's SFR via \autoref{eq:SFRvol}.

To solve the cubic equation for $H_{\rm g}$, we require values for $\Sigma_*$ and $\Sigma_{\rm g}$. The value of the stellar disk scale-height follows from $H_* = \Sigma_*/(2\rho_*)$, where $\rho_*$ has already been computed. We compute $\Sigma_*$ by walking the gravity tree within cylinders of fixed radius $500$~pc, oriented perpendicular to the galactic mid-plane. The total search height is set to $\pm 1$~kpc.  We compute $\Sigma_{\rm g}$ by walking the gas cell neighbor tree in the same way. We note that for $\Sg$, this may need to be adjusted for large cosmological simulations to exclude the influence of hot galactic winds that do not contribute to the gravitational potential acting on the gas disk. For each cylindrical tree-walk, we weight the particle masses using a 2D cubic spline kernel.

It is important to note that the gas cell neighbor tree in {\sc Arepo} is a separate structure from the gravity tree. {\sc Arepo} uses an efficient hierarchical time-stepping procedure, whereby the physical properties of particles are updated at time intervals $\Delta t_i = \Delta t_{\rm max} / 2^i$, where $\Delta t_{\rm max}$ is the maximum ``gravitational'' time-step for the simulation, and $i$ is an integer representing the time-step bin. Particles are assigned to bins according to their dynamical times. The full gravity/force tree of all particles is therefore only constructed at gravitational time-steps, at an interval of $\Delta t_{\rm max}$. On the other hand, the gas cell neighbor tree can be locally reconstructed on hierarchical time-steps. This means that the quantities $\rhod$, $\rhos$ and $\Ss$ can only be computed on gravitational time-steps during simulation run-time, whereas all other quantities in \autoref{eq:H_rho} can be computed on local hierarchical time-steps. Any changes to the depletion time between global time-steps are therefore due to iterative updates to $\Sigma_{\rm g}$, $\sigma_{\rm eff}$ and $P_{\rm eff}$, in practice.

We do not expect the lower frequency of updates to $\rhod$, $\rhos$ and $\Ss$ to have an adverse impact on the depletion time calculation. These three quantities enter only into the computation of the gravitational potential, and so will not evolve on time-scales shorter than a gravitational time-step.

To the extent that equilibrium holds, the pressure $P_\mathrm{eff}$ at the midplane is expected to agree with the weight obtained from \autoref{eq:weight_total}, which is equal to $\mathcal{W}=\seff^2\Sg/(2\Hg)$ when using the solution of \autoref{Eqn::Hg_cubic_resolved_Sigma} for $\Hg$. In practice, this means that since we are applying an EoS to obtain the pressure from density, the disk tends to adjust until the mid-plane $P_\mathrm{eff}$ and $\mathcal{W}$ match, as we test from the simulation. 
Given the equilibrium prediction for the gas scale height computed by solving the cubic in \autoref{Eqn::Hg_cubic_resolved_Sigma}, we can also obtain a prediction for the equilibrium mid-plane density $n_H=\Sg/(2\Hg m_H\mu)$.  The equilibrium prediction can be compared to the measured gas density.  


\subsection{Subgrid Model When Disk Scale-Heights Are Unresolved (\PRFMU)} \label{Sec::unresolved-case}
If both the gas and stellar disk scale-heights are unresolved in a simulation, as is typically the case at resolutions lower than $\sim10^5\Msun$, we can no longer assume that the values of the gas volume density or mid-plane pressure within the simulated gas reservoir are reliable. In the unresolved case, we must therefore 
solve for $H_{\rm g}$ from \autoref{eq:combined_Hg_eq} in terms of the gas and stellar surface densities $\Sigma_{\rm g}$ and $\Ss$, with the result (treating $\Hs/\Hg$ as a parameter) given in \autoref{eq:H_Sigma}.

If we do not explicitly separate the bulge and disk stellar contributions,
we can write the result as 
\begin{eqnarray}\label{eq:Hg_unresolved}
\Hg = \frac{2\seff^2}{\pi G[\Sg + \Sigma_*2/(1+\Hs/\Hg)]} \times \hskip1.2in \nonumber \\ 
\left(1 
+ \left[
1 +
 \frac{(16\pi/3) G \rho_{\rm d} \seff^2}{(\pi G)^2[\Sg + \Sigma_* 2/(1+\Hs/\Hg)]^2}\right]^{1/2}\right)^{-1},
 \hskip0.6cm
\end{eqnarray}
where $\Sigma_{\rm g}$ is computed by performing a tree-walk over the neighbor tree within a plane-perpendicular column of radius $500$~pc, as described in Section~\autoref{Sec::resolved-case}. $\Sigma_*$ is similarly computed by performing an identical cylindrical tree-walk over the gravity tree, with a kernel centered on the star-forming gas cell in question.  We shall adopt the formula in \autoref{eq:Hg_unresolved} for the equilibrium disk thickness $H_\mathrm{g}$ in the case that the mass resolution in the simulation is too coarse to directly resolve the gas disk's realistic thickness.  Given this, the equilibrium number density is defined as 
\begin{equation}\label{eq:n_H_equil_def}
n_\mathrm{H} = \frac{\Sg}{2 H_\mathrm{g} m_H \mu}.    
\end{equation}

In the case where stellar and gas scale heights are unresolved, we must make some assumption for their ratio in order to use \autoref{eq:Hg_unresolved}.  
One approach would be to simply assume that the stellar and gas disks have the same thickness, $\Hs/\Hg =1$, which is likely a reasonable approximation at high redshift.  However, as noted in \autoref{sec:parameters}, this could underestimate $\Hs/\Hg$ in the central regions of low-redshift galaxies.  Thus, for the present tests of \PRFMU, in which we consider a Milky-Way like galaxy, we shall adopt $\Hs/\Hg= \sigma_{*,z}/\seff$. The value of $\sigma_{*,z}$ is measured in the simulation by taking the set of gas cells used to compute $\Sigma_{*}$ and computing the cubic spline-weighted standard deviation of their vertical velocities. The value of $\seff$ is obtain from the subgrid EoS, assuming $P_\mathrm{eff} = \mathcal{W}$.  
From \autoref{eq:Hg_unresolved} we thus have 
\begin{eqnarray}\label{eq:tdyn_unresolved}
\tdyn = \frac{4\seff}{\pi G[\Sg + \Sigma_* 2/(1 + \sigma_{*,z}/\seff)]} \times \hskip1.2in \nonumber \\ 
\left(1 
+ \left[
1 +
 \frac{(16\pi/3) G \rhod\seff^2}{(\pi G)^2[\Sg + \Sigma_* 2/(1 + \sigma_{*,z}/\seff)]^2}\right]^{1/2}\right)^{-1}
 \hskip1cm
\end{eqnarray}
for the dynamical time $\tdyn = 2 \Hg/\seff$
and 
\begin{eqnarray}
\label{eq:P_unresolved}
\mathcal{W} = \frac{\pi G\Sg[\Sg + \Sigma_* 2/(1 + \sigma_{*,z}/\seff)]}{4} \times \hskip1.2in \nonumber \\ 
\left(1 
+ \left[
1 +
 \frac{(16\pi/3) G \rho_{\rm d} \seff^2}{(\pi G)^2[\Sg + \Sigma_*2/(1 + \sigma_{*,z}/\seff)]^2}\right]^{1/2}\right)
 \hskip1cm 
\end{eqnarray}
for the predicted equilibrium weight of the ISM disk in the total gravitational potential.  

We then use \autoref{eq:P_unresolved} to set the value of $P_{\rm eff} \rightarrow {\cal W} $, and thus to set the values of $\sigma_{\rm eff}$ from \autoref{Eqn::sigma_eff}. Note that because $H_{\rm g}$ and $\cal W$ depend on $\sigma_{\rm eff}$ and vice versa, the values obtained from \autoref{eq:P_unresolved} and \autoref{Eqn::sigma_eff} must be iterated a handful of times, until convergence is achieved.  After convergence is achieved, we use the resulting $P_{\rm eff} \rightarrow {\cal W} $ in \autoref{Eqn::upsilon} for $\Upstot$.  

As in the \PRFMR\ implementation, the values of $\sigma_{\rm eff}$, $\Upsilon_\mathrm{tot}$ and $\tdyn$ can then be used to set the depletion time via \autoref{Eqn::tdep_general}. When using \autoref{eq:tdyn_unresolved}, we obtain
\begin{eqnarray}\label{eq:tdep_unresolved}
\tdep = \frac{4\Upsilon_\mathrm{tot}}{\pi G[\Sg + \Sigma_*2/(1 + \sigma_{*,z}/\seff) ]} \times \hskip1.2in \nonumber \\ 
\left(1 
+ \left[
1 +
 \frac{(16\pi/3) G \rho_{\rm d} \seff^2}{(\pi G)^2[\Sg + \Sigma_*2/(1 + \sigma_{*,z}/\seff)]^2}\right]^{1/2}\right)^{-1}.
  \hskip1cm 
\end{eqnarray}
As explained in \autoref{sec:volumetric}, no renormalization is needed relative to the patch-averaged calibrations for the \PRFMU\ implementation. 

We note that in the case where dark matter is negligible, \autoref{eq:P_unresolved} simplifies to 
${\cal W}\rightarrow \pi G \Sg[\Sg +\Ss 2/(1 + \sigma_{*,z}/\seff)]/2$ and \autoref{eq:tdep_unresolved} simplifies to $\tdep = (2 \Upsilon_\mathrm{tot}/\pi G)/[\Sg +\Ss2/(1 + \sigma_{*,z}/\seff)]$.  In the case where $\Hs/\Hg=\sigma_{*,z}/\seff=1$ is adopted, this would further simplify to   
${\cal W}\rightarrow \pi G \Sg[\Sg +\Ss ]/2$ and  $\tdep = (2 \Upsilon_\mathrm{tot}/\pi G)/[\Sg +\Ss]$.  In this situation, there would be no explicit dependence of $\tdep$ (and therefore of the SFR) on the EoS, with only $\Sg$ and $\Ss$ entering in determining $\tdep$.

We emphasize that in the \PRFMU\ implementation, the gas density and stellar density on the Voronoi mesh are not directly used in computing $\tdep$. Only $\Sg$ and $\Sg$, which are vertically integrated quantities and therefore may be reliably evaluated provided the disk is \textit{radially} resolved, are used.  Furthermore, while here we shall adopt the EoS calibrated from TIGRESS for the cosmological simulation, in principle any prescription for pressure that stabilizes the numerical solution may be adopted, because the density and pressure are not directly used in computing the star formation.  That is, in some cases it may be numerically necessary to adopt a stiffer EoS (such as that in TNG) in order prevent fragmentation of a disk with high surface density, since in the absence of explicit feedback there is nothing to disperse bound structures if they form.  However, in using \autoref{eq:tdep_unresolved} for $\tdep$ in individual cells, the value of $\seff$ may still be set based on the calibration of resolved ISM simulations, i.e. using  \autoref{eq:P_unresolved} for $P_\mathrm{eff}\rightarrow\mathcal{W}$ in \autoref{Eqn::sigma_eff} and iterating.

\section{Simulations and Results}\label{sec:mod_results}
In this section, after briefly describing the initial conditions of our simulations (\autoref{sec:init}), 
we begin by validating the key components of the \PRFMR\  and \PRFMU\ implementations (\autoref{sec:valid}): we  
demonstrate that both implementations properly capture the relationships, as  calibrated using the TIGRESS simulations, between (1) density and pressure (or weight), and (2) dynamical time and depletion time. 
Here, we also show how the TIGRESS EoS and SFR model prescriptions differ from those adopted in the \TNG\ simulations.

We then present our main results (\autoref{sec:model_comparison}), comparing the SFRs and timescales between simulations with the \PRFMR\ and \PRFMU\ implementation at mass resolution of 10$^5$, 10$^6$, and 10$^7 \Msun$.  We shall show that the \PRFMU\ simulations at all resolutions are able to recover star formation history and distributions as obtained from the \PRFMR\ simulation at resolution of 10$^5 \Msun$.  We also 
demonstrate the common failing of traditional approaches to setting SFRs solely based on volume density: when resolution is too coarse, star formation is inconsistent with what is obtained at high resolution. This lack of robustness to resolution changes afflicts simulations adopting both the  \PRFMR\ and \TNG\ prescriptions at low resolution.  Since the \PRFMU\ SFR prescription employs surface densities rather than volume densities, it does not suffer from this defect. 

Finally, we analyze the properties of both the \PRFMR\ (\autoref{Sec::PRFMR-props}) and \PRFMU\ (\autoref{Sec::PRFMU-sims}) simulations in more detail.  This analysis reveals what is behind the lack of robustness of traditional ``volumetric'' SFR approaches to coarsening of numerical resolution.  It also shows why the \PRFMU\ approach offers the promise of  recovering realistic distributions of star formation even at the low numerical resolutions that are affordable for large-box cosmological simulations.   

For many of our comparisons, we use results from our simulations at a common evolutionary time of 50 Myr; this is after early transients have settled out and before large-scale galaxy properties significantly diverge from each other due to  differences in subgrid prescriptions or resolution.

\subsection{Initial conditions}\label{sec:init}

All initial conditions used in this work are based on the Agora suite of isolated galaxy simulations \citep{Kim14}. The initial condition at $10^5\Msun$ resolution is a Gadget-format initial condition generated by the Agora collaboration, and the initial conditions at $10^6\Msun$ and $10^7\Msun$ are copies at lower resolution, generated using {\sc MakeNewDisk}. The Agora disk is designed to resemble a Milky Way-like galaxy at redshift $z=0$, with an NFW dark matter halo of mass $M_{200} = 1.07 \times 10^{12} \Msun$, halo concentration parameter $c=10$ and spin parameter $\lambda = 0.04$, an exponential disk of mass $4.297 \times 10^{10}\Msun$, scale-length of $3.43$~kpc, initial scale-height 343~pc, and gas fraction 0.18, and a Hernquist-type bulge of mass $3.437 \times 10^9\Msun$.

\begin{figure*}
	\includegraphics[width=\linewidth]{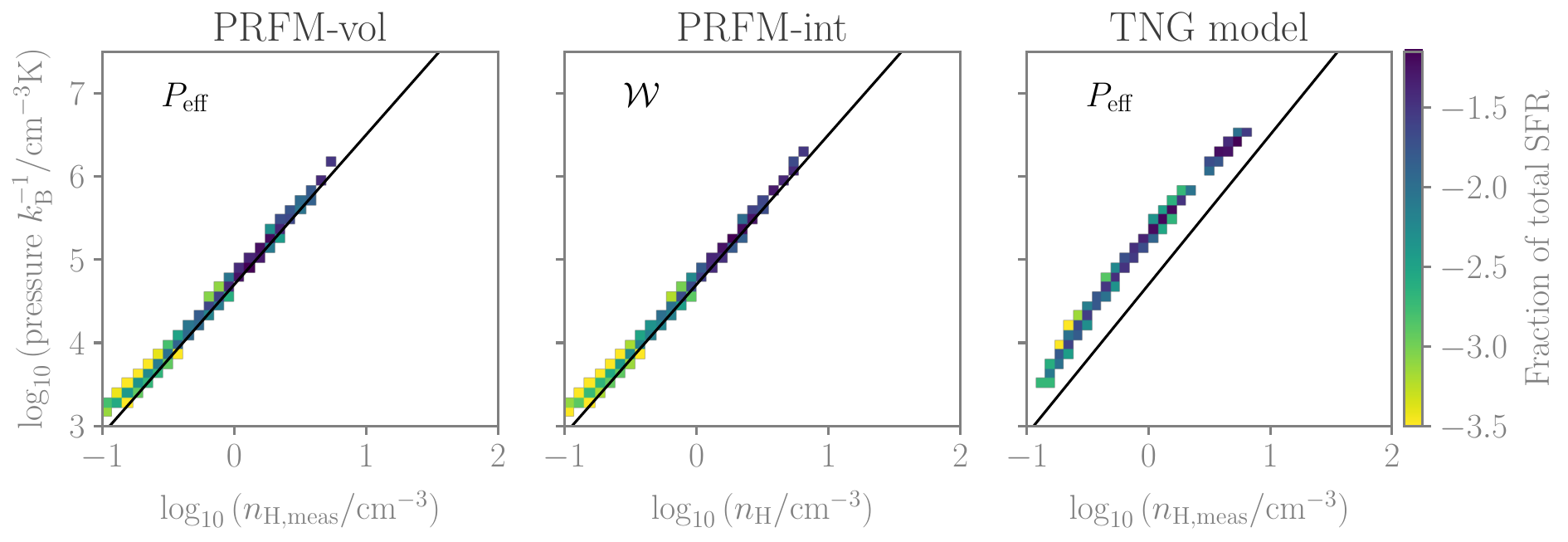}
	\caption{Equation of state comparison.  Distribution of the gas effective pressure $P_{\rm eff}$ (left and right) or weight $\mathcal{W}$ (center) as a function of the measured gas hydrogen number density $n_{\rm H,meas}$ (left and right) or computed equilibrium density $n_H$ (center),  in pixels of scale $1 \times 1$~kpc$^2$, and averaged over all simulation time-stamps with an interval of $10$~Myr between $50$~Myr and $600$~Myr, inclusive. All results are from simulations with mass resolution $10^5\Msun$. The thick black line represents the best-fit to the TIGRESS simulations (see \autoref{Eqn::prfm-eos}).}\label{fig:EoS}
\end{figure*}
\begin{figure*}
	\includegraphics[width=\linewidth]{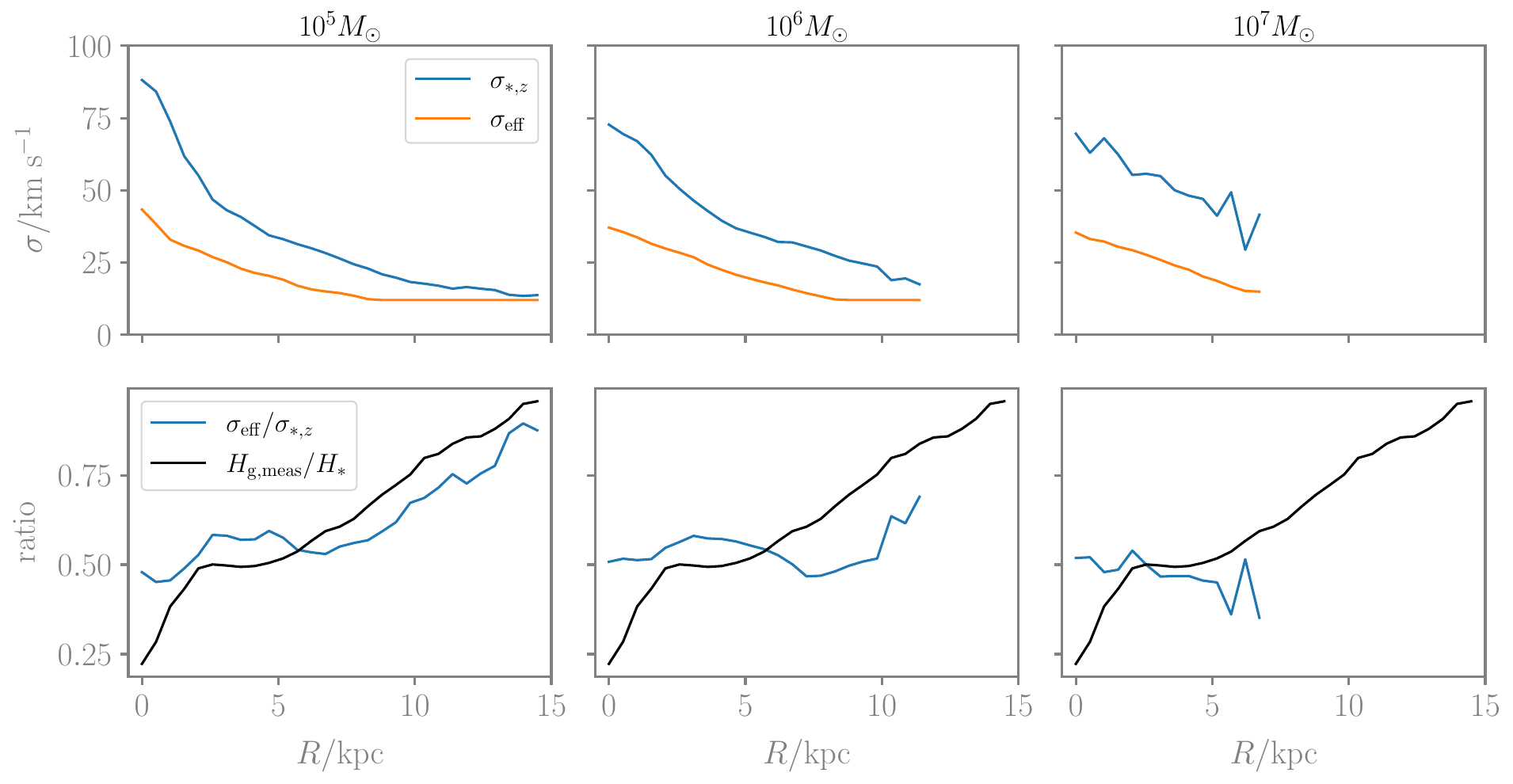}
	\caption{{\rm Upper row:} Effective gas velocity dispersion ($\sigma_{\rm eff}$, orange, computed from the effective EoS), and measured stellar velocity dispersion ($\sigma_{*,z}$, blue) in the vertical direction as a function of galactocentric radius $R$ from simulations adopting the \PRFMU\ implementation, at resolutions of $10^5$, $10^6$ and $10^7\Msun$. {\rm Lower row:} Ratio of these two velocity dispersions (blue lines), in comparison to the measured ratio of the gas and stellar scale-heights $H_{\rm g,meas}/H_*$ in the \PRFMR\ implementation at $10^5\Msun$ resolution (black lines). }
    \label{Fig::veldisp-ratio}
\end{figure*}

\subsection{Validation of implementation}\label{sec:valid}
First, we confirm our numerical implementation and model prescription. \autoref{fig:EoS} shows, for the \PRFMR, \PRFMU, and \TNG~simulations, the EoS, i.e. the relationship between pressure and density. The solid black line represents the expected relationship calibrated in the TIGRESS simulations and set via \autoref{Eqn::prfm-eos} and~\autoref{Eqn::sigma_eff} in our simulations.  For all panels, results  are from simulations with mass resolution $10^5\Msun$, based on snapshots between times of $50$~Myr and $600$~Myr, inclusive.  

In the left-hand panel we show the actual EoS in the \PRFMR\ simulation, relating individual gas cell densities $n_{\rm H, meas}$ and gas cell pressures $P_{\rm eff}$ which are set via \autoref{Eqn::prfm-eos}. This EoS is also adopted here for the \PRFMU\ implementation, but for this simulation (center panel) we prefer to show the relationship between the computed mid-plane density $n_{\rm H}$ (\autoref{eq:n_H_equil_def}) and the computed ISM weight $\mathcal{W}$ (\autoref{eq:P_unresolved}), as it is $\W$ that is used to set the feedback yield $\Upstot$, the effective velocity dispersion $\seff$, and ultimately
$t_{\rm dep}$ in the \PRFMU\ implementation. Finally, in the right-hand panel, we show the EoS adopted for the \TNG\ simulation, in comparison to that set from the  TIGRESS calibration. We see that the \TNG\ prescription has a higher pressure relative to density than is predicted by the  TIGRESS multiphase ISM simulations \citep{Burger2025}.

Second, we consider the ratio of velocity dispersions $\sigma_{\rm eff}/\sigma_{*,z}$, used as a proxy for the ratio of gas to stellar scale-heights $H_{\rm g, meas}/H_*$ in the \PRFMU\ implementation (in both \autoref{eq:P_unresolved} and \autoref{eq:tdep_unresolved}).  In \autoref{Fig::veldisp-ratio} the upper panels show the effective gas (orange lines) and stellar (blue lines) vertical velocity dispersions individually, demonstrating that they are each reasonably well-converged with resolution. The lower panels show that their ratio (blue lines) is a good proxy for the ratio of scale-heights (black lines) in the \PRFMR\ simulation at $10^5\Msun$ resolution, which may be computed directly from the gas and stellar mid-plane densities and surface densities. \autoref{Fig::veldisp-ratio} therefore demonstrates that $\sigma_{\rm eff}/\sigma_*$ can reliably be used in instances for which the true scale-heights are not resolved.

\begin{figure*}
	\includegraphics[width=\linewidth]{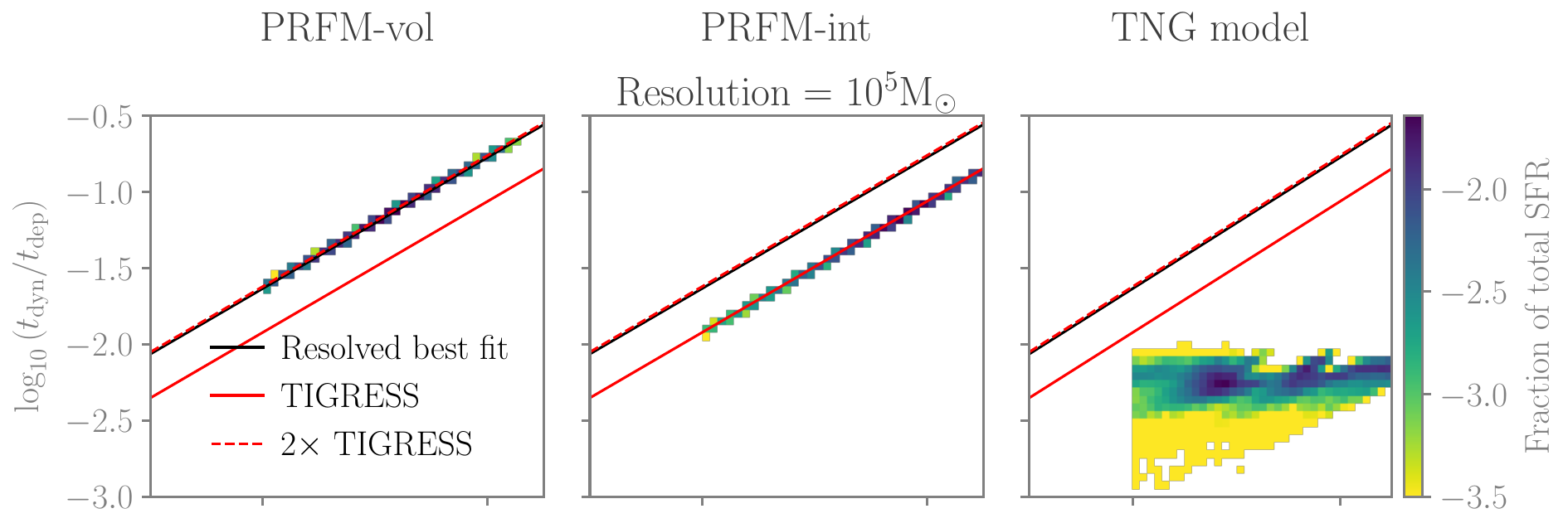}
    \includegraphics[width=\linewidth]{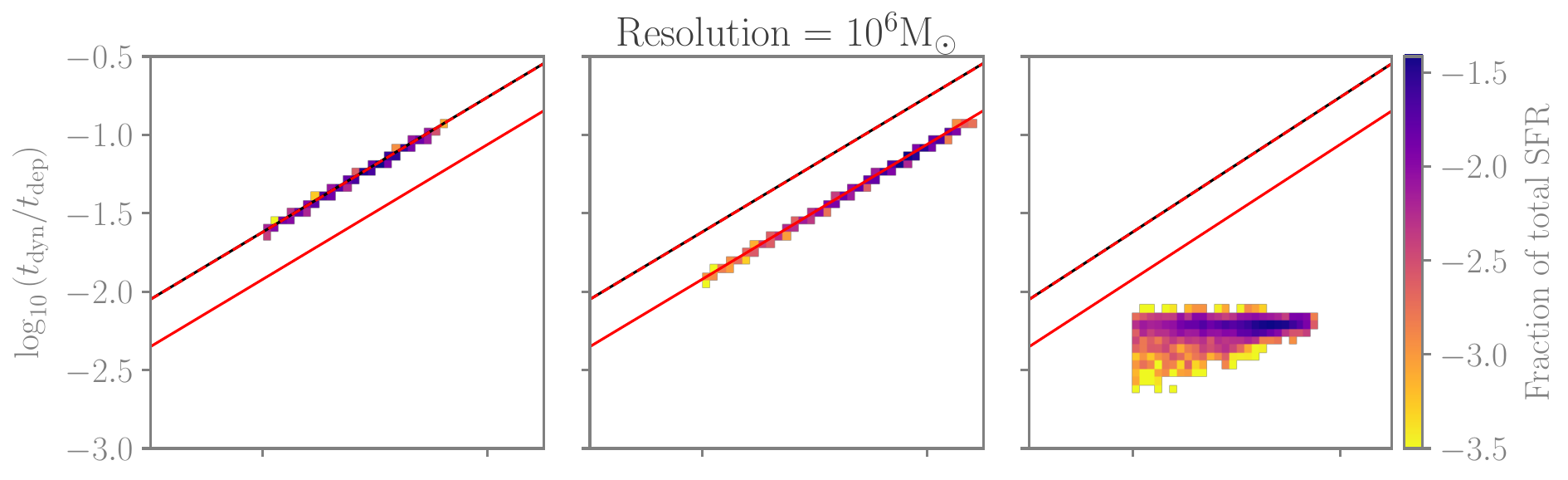}
    \includegraphics[width=\linewidth]{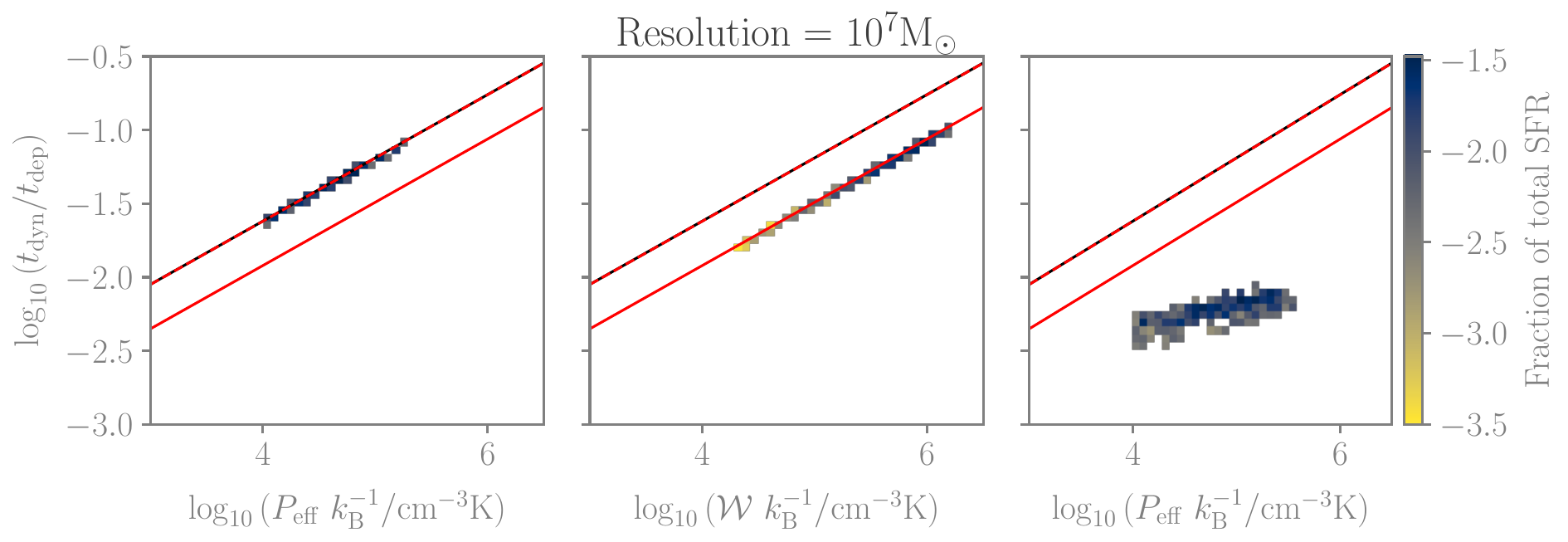}
	\caption{Ratio of dynamical time to depletion time, i.e. the star formation efficiency per dynamical time $\varepsilon_\mathrm{dyn}=\tdyn/\tdep$ (see \autoref{eq:SFR}), as a function of gas pressure or weight. We show the distribution of gas cells in the plane spanned by the measured mid-plane effective pressure $P_{\rm eff}$ (left and right columns) or computed weight $\mathcal{W}$ (center column)  and the ratio of the dynamical time to depletion time $t_{\rm dyn}/t_{\rm dep}$. All three simulation resolutions of $10^5\Msun$ (top row), $10^6\Msun$ (center row) and $10^7\Msun$ (bottom row) are compared at a simulation time of $50$~Myr. The lower (solid) red lines represent the original TIGRESS fit, i.e. \autoref{eq:SFRvol} with $\mathcal{R}_f=1$, and the upper (dashed) red line represents \autoref{eq:SFRvol} with $\mathcal{R}_f=2$,  based on  \autoref{eq:renorm_fac}. The black lines represent the least-squares best fit ($\log{t_{\rm dyn}/t_{\rm dep}} \approx 0.43 \log{P} - 3.34$) to the \PRFMR\  simulation at $10^5\Msun$, including all gas cells with ${\rm SFR} > 0$. We note that the black lines overlap completely with the upper (dashed) red line, as intended.}
    \label{Fig::1e5_taurat}
\end{figure*}

Our third point of validation is presented in \autoref{Fig::1e5_taurat}, in which we demonstrate that the \PRFMR\ and \PRFMU\ implementations obey the relationship between pressure (x-axis) and the ratio of dynamical time to depletion time ($t_{\rm dyn}/t_{\rm dep}\equiv \varepsilon_\mathrm{dyn}$, y-axis) set out in \autoref{eq:SFRvol}. The red solid (lower) diagonal lines in each panel represent the original TIGRESS prediction for the ratio of mid-plane dynamical time $t_{\rm dyn}$ to vertically-averaged depletion time $t_{\rm dep}$, as a function of the ISM midplane weight or pressure $\W=\Peff$.  As a volumetric SFR, this is \autoref{eq:SFRvol} with $\mathcal{R}_f=1$, which is appropriate for a vertically unresolved system. The red dashed (upper) diagonal lines in each panel represent the renormalization of the TIGRESS calibration appropriate for a fully vertically resolved disk, which is \autoref{eq:SFRvol} with $\mathcal{R}_f=2$. This renormalization factor accounts for the vertical variation of the volume density of star-forming gas in each column of gas, following \autoref{eq:renorm_fac}. For the first column of \autoref{Fig::1e5_taurat}, we use $P_{\rm eff}$ to evaluate the second line of \autoref{eq:SFRvol}, as appropriate for the \PRFMR\ implementation.  For the second column  of \autoref{Fig::1e5_taurat}, we use $\W$ to evaluate the second line of \autoref{eq:SFRvol}, as appropriate for the \PRFMU\ implementation.  For the third column, we use the \TNG\ prescription for $\tdep$ and compute $\tdyn$  using local densities in \autoref{eq:tdynrho}.

For both \autoref{Fig::1e5_taurat}, and subsequent figures, we use distinct color maps for simulations at different mass resolutions ($10^5\Msun$, $10^6\Msun$, and $10^7\Msun$), and the color scale is used to show the fraction of the contribution to the total SFR from each pixel in the two-dimensional histogram. 
We can see that the time-scale ratios in both the \PRFMR\ and \PRFMU\ cases, at all mass resolutions, agree well with the appropriate prescriptions based on TIGRESS calibrations. This is expected, as \autoref{eq:SFRvol} is used directly in {\sc Arepo} to set the depletion time as a function of the dynamical time. As is evident from the right column, we also note that the TNG implementation leads to a time-scale ratio with significantly longer depletion times relative to the dynamical time (i.e., lower star formation efficiency). 

This figure highlights an important feature of the PRFM model in comparison to traditional star formation prescriptions, such as that adopted in TNG.  Namely, not only is the star formation efficiency $\varepsilon_\mathrm{dyn}$ larger for the PRFM model than for TNG, but also $\varepsilon_\mathrm{dyn}$ increases significantly in environments where the density and pressure increase, whereas in traditional prescriptions  $\varepsilon_\mathrm{dyn}$ is constant. Figure 9 of \citet{Hassan2024} showed that this can make a significant difference to expectations for star formation, especially 
under conditions present at higher redshifts.
We will discuss this difference between the TNG and PRFM prescriptions at greater length in the following subsections.

\begin{figure*}
	\includegraphics[width=\linewidth]{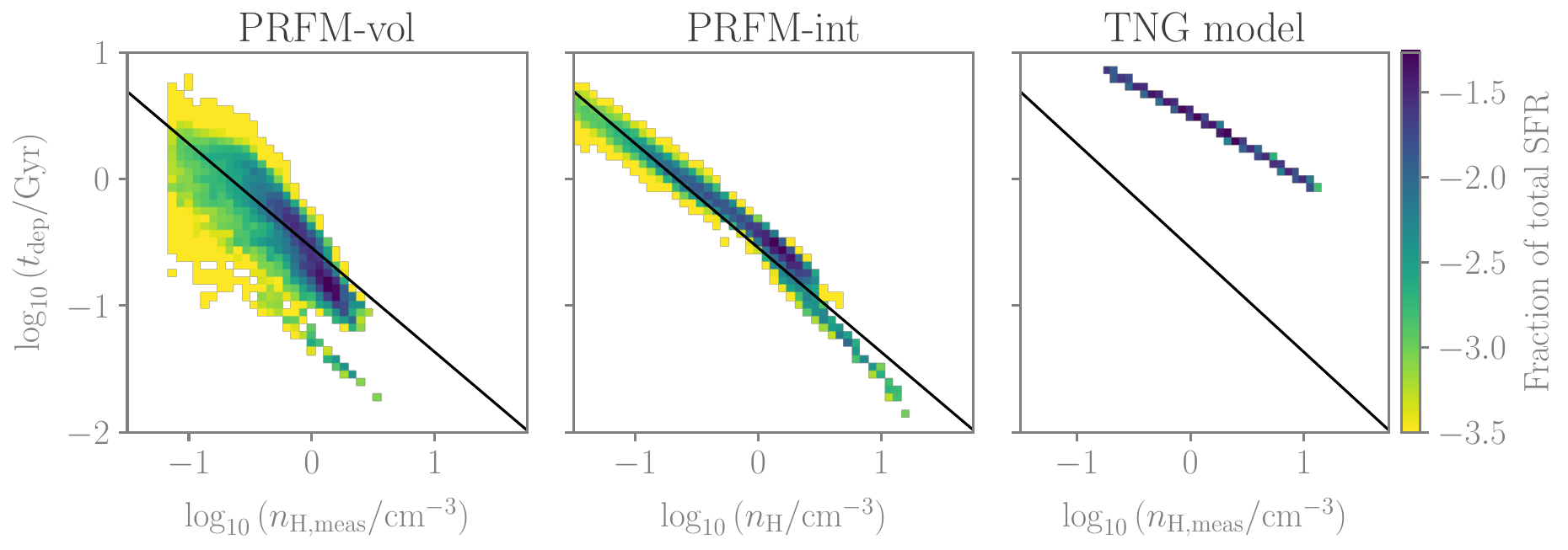}
	\caption{Depletion time as a function of gas density, from simulations with mass resolution $10^5\Msun$, at a simulation time of $50$~Myr. The gas cell depletion time is $t_{\rm dep} = M_{\rm g}/\dot{M}_*$, and  density is measured in the simulation as $n_\mathrm{H,meas}$ (left and right)  or computed as an equilibrium value $n_\mathrm{H}$ (center).   The black lines represent the least-squares best fit ($\log{t_{\rm dep}} \approx -0.96 \log{n_{\rm H}} - 0.46$) to the \PRFMR\ simulation, including all gas cells with ${\rm SFR} > 0$.}
    \label{Fig::SFR_density_tauDep}
\end{figure*}

\begin{figure*}
	\includegraphics[width=\linewidth]{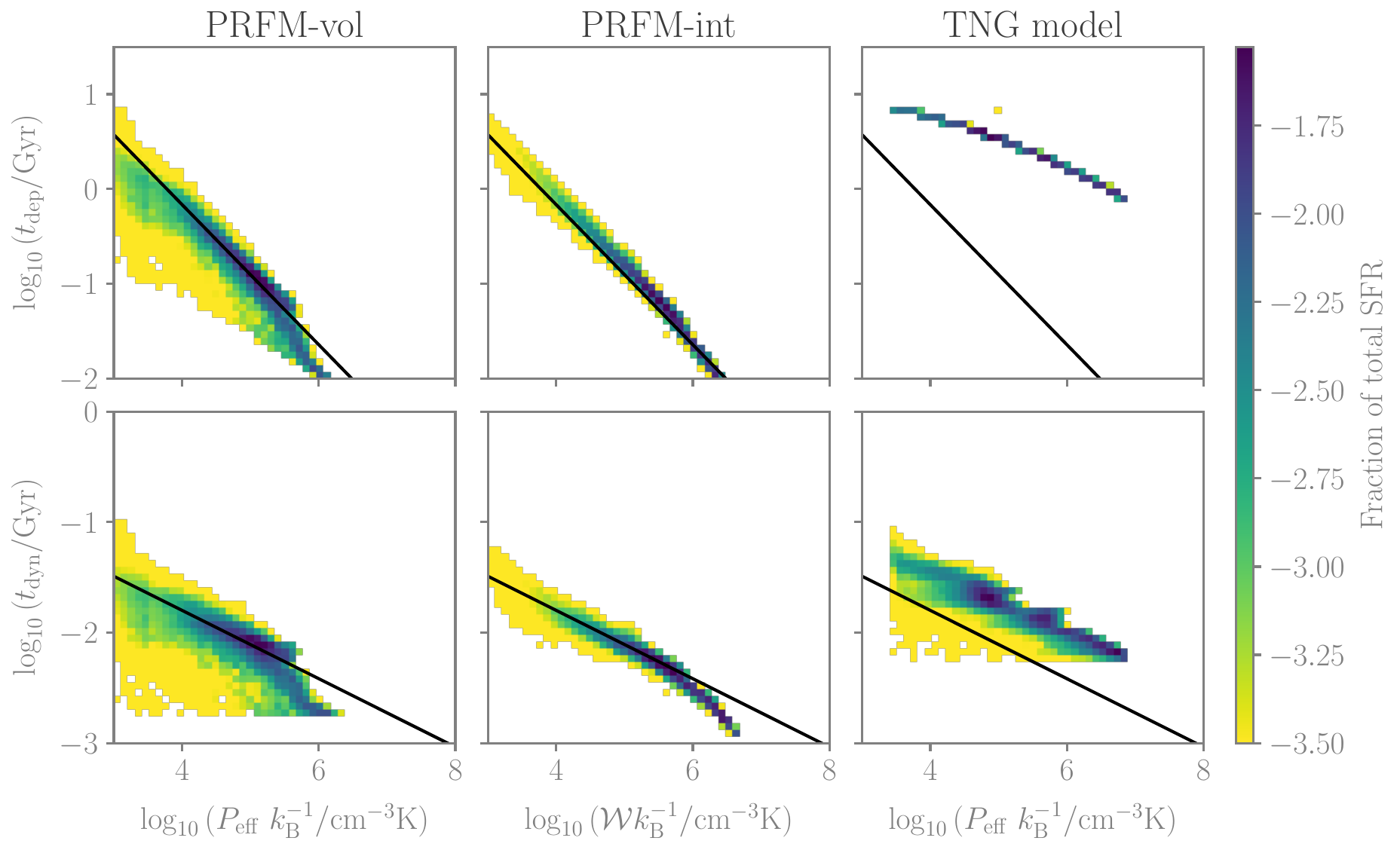}
	\caption{Depletion and dynamical time as a function of gas pressure, from simulations with mass resolution $10^5\Msun$ at time 50 Myr. We show the distribution of gas cells in the plane spanned by the gas cell measured effective pressure $P_{\rm eff}$ (left and right columns) or computed weight $\mathcal{W}$ (center column) and the gas cell depletion time $t_{\rm dep}\equiv M_\mathrm{g}/\dot{M}_*$ (top row) or the gas cell dynamical time $t_{\rm dyn}$ (bottom row).   
    For \PRFMR\ and \TNG\ simulations (left and right), $t_\mathrm{dyn}$ is computed via~\autoref{eq:tdynrho}; for \PRFMU\ (center),  $t_\mathrm{dyn}$ is computed via~\autoref{eq:tdyn_unresolved}. The black lines represent the least-squares best fits ($\log{t_{\rm dep}} \approx -0.62 \log{P_{\rm eff}} + 2.52$ and $\log{t_{\rm dyn}} \approx -0.27 \log{P_{\rm eff}} - 0.76$) to the \PRFMR\ simulation, including all gas cells with ${\rm SFR} > 0$.}
    \label{fig:P_vs_tdep_tdyn}
\end{figure*}

Finally, in \autoref{Fig::SFR_density_tauDep} and \autoref{fig:P_vs_tdep_tdyn}, we check whether the \PRFMR~and \PRFMU~implementations produce similar values of the depletion time at $10^5\Msun$ resolution, 
and compare these depletion times to those produced in the \TNG~implementation.

In \autoref{Fig::SFR_density_tauDep} we show the depletion time $t_{\rm dep}$ as a function of the directly-measured mid-plane density $n_{\rm H, meas}$ in the \PRFMR~and \TNG~(left and right columns), and as a function of the derived mid-plane gas density $n_{\rm H} = \Sigma_{\rm g} /( 2 H_{\rm g} n_{\rm H} \mu)$ in \PRFMU~(center column). The solid black lines represent the least-squares best fit to the gas cell distribution in \PRFMR. Similarly, in \autoref{fig:P_vs_tdep_tdyn}, we show the depletion time as a function of the directly-measured effective mid-plane pressure $P_{\rm eff}$ in the \PRFMR~and \TNG, and as a function of the derived ISM weight $\mathcal{W}$ in \PRFMU. Again, the solid black lines represent the least-squares best fit to the gas cell distribution in \PRFMR.

These comparisons demonstrate good agreement between the depletion times produced using \PRFMR\ and \PRFMU\ in $10^5\Msun$ resolution simulations, with the gas cell distribution for \PRFMU\ falling on or close to the best-fit lines generated by \PRFMR. The main difference between the two implementations is in the spread of depletion times seen at a given density or pressure. This is to be expected, considering \PRFMR\ assigns a different depletion time within each gas cell, whereas \PRFMU~assigns the same depletion time to all gas cells within the vertical columns used to compute $\Sigma_{\rm g}$ and $\mathcal{W}$. 

Both \PRFMR\ and \PRFMU\ display an offset between disk- and bulge-dominated regions, in which the depletion time drops due to a sharp increase in stellar surface density $\Sigma_*$. 
This can be seen in both figures, as a dip below the best-fit line at high densities and pressures.
Because $\tdep$ depends not just on the gas density but also on the stellar density, the dark matter density, and the pressure (see \autoref{eq:SFRvol_den}), for \PRFMR\ there is a tighter correlation between $\tdep$ and $\Peff$ in \autoref{fig:P_vs_tdep_tdyn} than the correlation between  $\tdep$ and $n_{\rm H, meas}$ in \autoref{Fig::SFR_density_tauDep}. 

Finally, in \autoref{fig:P_vs_tdep_tdyn} we also show $\tdyn$ as a function of $\Peff$ or $\W$. This demonstrates the agreement between $\tdyn$ in \PRFMU\ and \PRFMR\ at $10^5\Msun$ resolution.  In addition, this shows that $\tdyn$ is generally longer in the \TNG\ simulation, which directly reflects the larger coefficient adopted in the EoS.

We note that the depletion times prescribed in the PRFM implementations vary over a substantially larger range than those prescribed in TNG, and reach substantially lower values by up to two orders of magnitude. Both the larger range and lower minimum values of $\tdep$ in PRFM are due to the increase in star formation efficiency per dynamical time at higher pressure (following the TIGRESS calibration in \autoref{eq:epsdyn_calib}). The generalized dynamical time itself also has a greater range in PRFM than in TNG, as it depends on the stellar as well as the gas density. By contrast, TNG assigns a constant value to the star formation efficiency per free-fall time, where this timescale depends only on gas density.

\begin{figure*}
	\includegraphics[width=\linewidth]{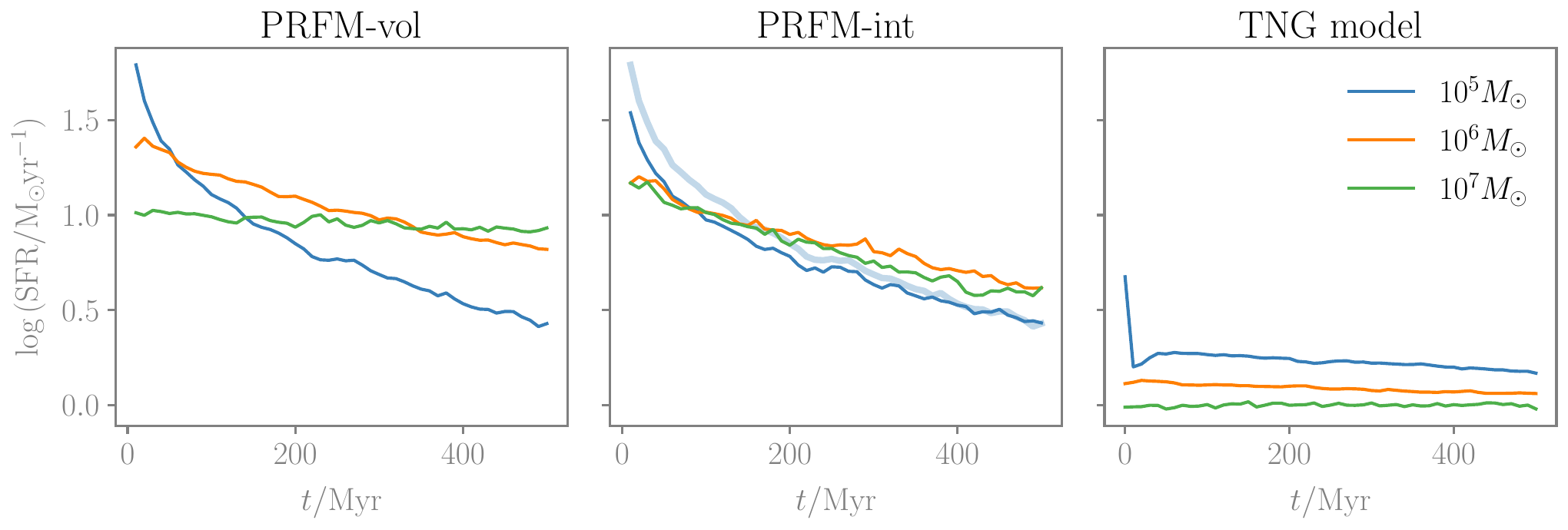}
	\caption{Time evolution of the total galactic SFR in each set of simulations (\PRFMR, \PRFMU, and \TNG\ implementations) at each of the three resolutions tested ($10^5\Msun$ in blue, $10^6\Msun$ in orange, and $10^7\Msun$ in green). The transparent blue line in the center panel is a copy of the $10^5\Msun$ evolution from the \PRFMR\ case (blue line, left panel).}
    \label{Fig::SFR_vs_time}
\end{figure*}

\begin{figure*}
	\includegraphics[width=\linewidth]{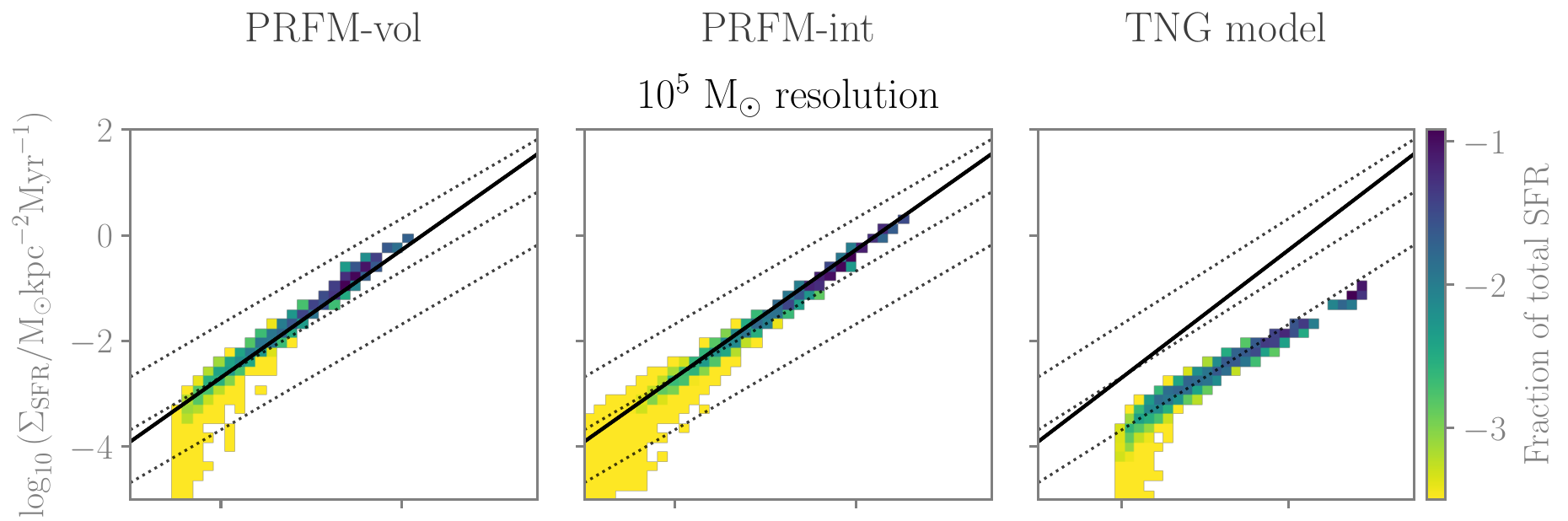}
    \includegraphics[width=\linewidth]{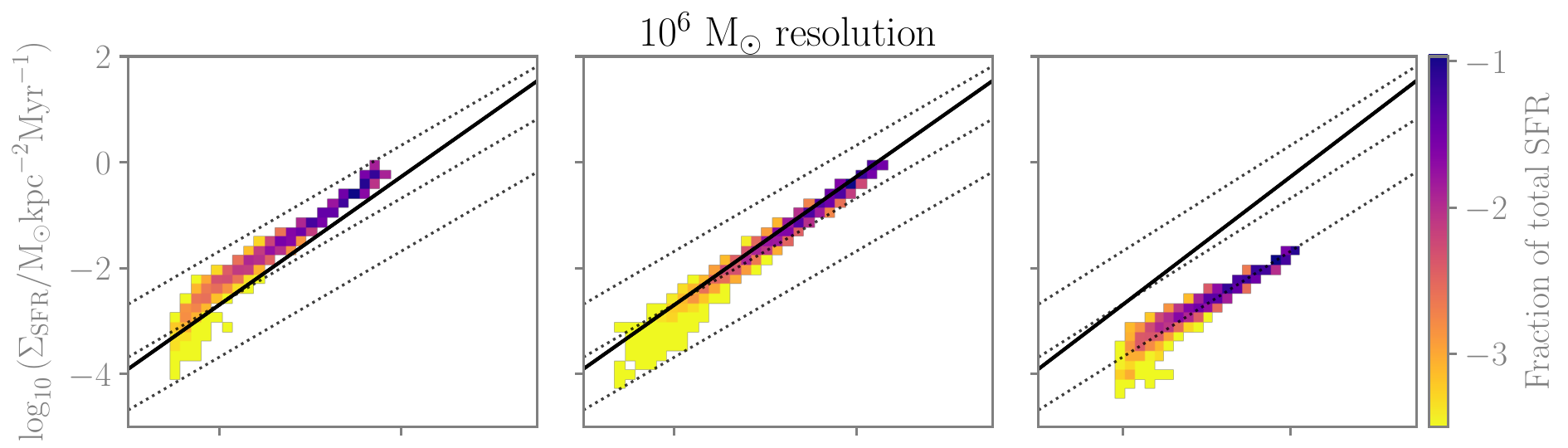}
    \includegraphics[width=\linewidth]{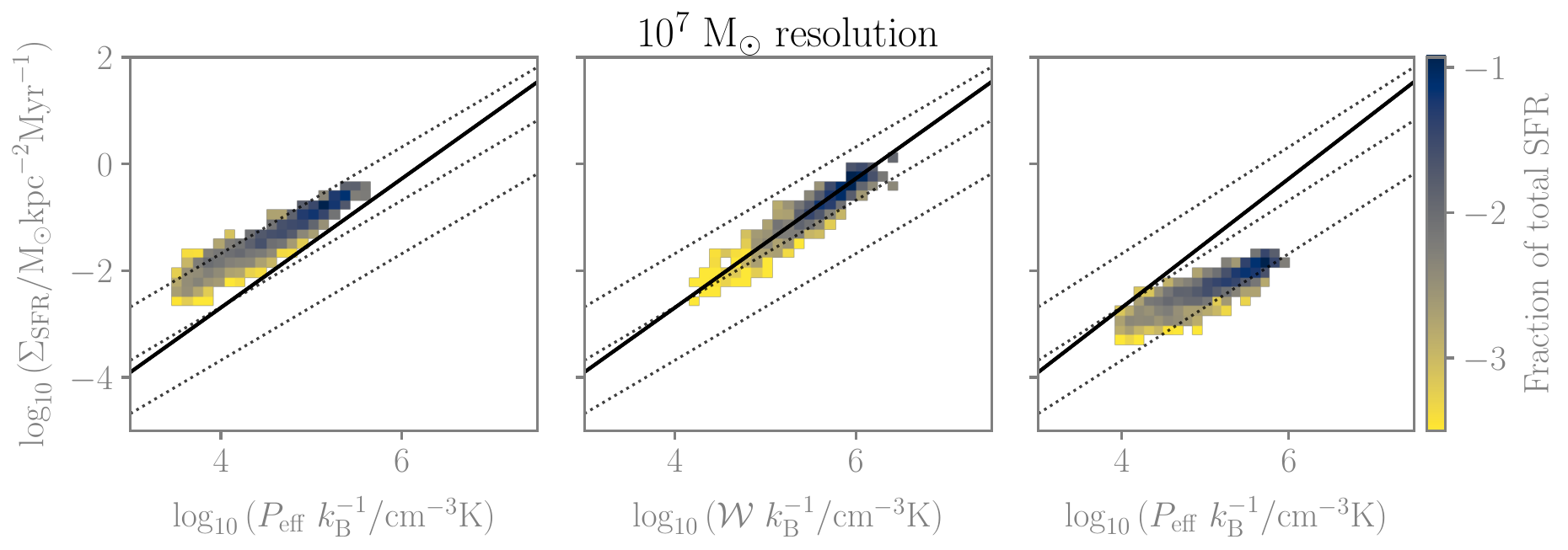}
	\caption{Distribution of SFR surface densities $\Sigma_{\rm SFR}$ as a function of the mid-plane effective gas pressure $P_{\rm eff}$ (left and right), or computed weight $\mathcal{W}$ (center),  measured in projected pixels of scale $1 \times 1$~kpc$^2$, and averaged over all simulation time-stamps between $50$~Myr and $600$~Myr, inclusive. 
    All three simulation resolutions of $10^5\Msun$ (top row), $10^6\Msun$ (center row) and $10^7\Msun$ (bottom row) are shown. 
    The thick black lines represent the best-fit to the TIGRESS simulations (see \autoref{Eqn::upsilon}), while the dotted lines from bottom to top are for the constant feedback yield of $10^4$, $10^3$, and $10^2$ $\kms$.}\label{fig:SSFR-P}
\end{figure*}

\subsection{Comparison of star formation rates and timescales in \PRFMR\ and \PRFMU}\label{sec:model_comparison}
In \autoref{Fig::SFR_vs_time}, we present a comparison of global SFRs obtained from simulations adopting  the \PRFMR\  (left), \PRFMU\ (center) and \TNG\ (right) implementations of the EoS and depletion time. The peaks in SFR at early times, in the \PRFMR\ and \PRFMU\ cases, are artificial, and are due to the initial collapse and equilibration of the isolated disk galaxies\footnote{The same initial condition is used at each resolution for all EoS and star formation implementations, which has an EoS close to the EoS used in the \TNG\ implementation. Because at the same density, this pressure is higher than that produced by the PRFM EoS, an initial phase of collapse and resettling occurs in the PRFM simulations, leading to the transient star formation `burst.'}.

Following the initial starburst, we see that the PRFM galaxies settle to a higher SFR than do the galaxies governed by the TNG EoS and star formation law, consistent with the lower depletion times reached within dense, high-pressure gas in the PRFM prescription (see \autoref{Fig::SFR_density_tauDep} and \autoref{fig:P_vs_tdep_tdyn}). We also see that the \PRFMU\  implementation offers a greater degree of resolution robustness than does the \PRFMR\ or \TNG\ implementation, with good agreement between the global SFR evolution across all resolutions for \PRFMU. The difference across resolution for  the  \PRFMR\ and \TNG\ implementations is to be expected, given that they rely on the volume densities within the disk, which become inaccurate when the disk scale height becomes spatially poorly-resolved or unresolved as the mass resolution is coarsened.

The improved resolution robustness of the \PRFMU\  over the \PRFMR\ implementation is also manifest in \autoref{fig:SSFR-P}, which shows the Ostriker-Kim relationship between the mid-plane pressure or weight (x-axis) and SFR surface density $\Sigma_{\rm SFR}$ (y-axis) for each implementation and at each resolution. For each implementation, $\Sigma_{\rm SFR}$ is compared to the pressure that directly sets the depletion time ($P_{\rm eff}$ for the \PRFMR\ and \TNG\ implementations, and $\mathcal{W}$ for the \PRFMU\ implementation). All quantities are measured in projected pixels of scale $1 \times 1$~kpc$^2$, and are averaged over all simulation time-stamps between 50 and 600~Myr.

Each row of the figure (with distinct colormaps) compares the three different implementations at a different resolution. For the $10^5\Msun$ resolution, both the \PRFMR\ and \PRFMU\ simulations closely follow the midplane pressure vs. $\SSFR$ relation calibrated in 
TIGRESS. However, for the two lower resolutions ($10^6$ and $10^7\Msun$), for \PRFMR\ we see an elevation in $\Sigma_{\rm SFR}$ at a given pressure, similar to the globally elevated SFR at low resolution in \autoref{Fig::SFR_vs_time}. By contrast, the \PRFMU\  implementation maintains good agreement with the relation calibrated in TIGRESS (black solid lines) across all resolutions.

The top row of \autoref{fig:SSFR-P} also demonstrates again that the higher and non-constant star formation efficiency in 
the PRFM model produces a marked difference in the SFRs obtained at high pressure (and density), relative to the implementation used in the \TNG\ simulations. In particular, the PRFM simulations have a steeper-than-linear relation between $\SSFR$ and midplane pressure (consistent with the calibration in TIGRESS), while this relation is significantly shallower in all of the \TNG\ simulations.  For reference, dotted lines in \autoref{fig:SSFR-P} indicate what would be linear relationships between pressure and $\SSFR$, i.e. constant feedback yield $\Upstot$ in \autoref{eq:P_SFR}, rather than $\Upstot$ that decreases at higher pressure as in \autoref{Eqn::upsilon}.  The bottom row of \autoref{fig:SSFR-P} shows that $\SSFR$ vs. pressure in \TNG\ simulations at low resolution do not agree with their higher-resolution counterparts. 

We note that more extensive comparisons of $\SSFR$ vs. $\W$, as well as $\SSFR$ vs. $\Sg$, are shown in Figures 11 and 12 of \citet{Hassan2024}.  There, we apply the PRFM model in post-processing to TNG50 galaxies 
at both $z=0$ and $z=2$, covering a wide range of galactic conditions, and compare to 
observed relations from PHANGS and EDGE nearby galaxy surveys. In this context, it is worth noting that the slope of observed star formation relations depend strongly on the adopted conversion factor $\alpha_\mathrm{CO}$ from CO intensity to molecular surface density, with variable $\alpha_\mathrm{CO}$ 
giving a steeper Kennicutt-Schmidt slope \citep{Narayanan12}. When variable $\alpha_\mathrm{CO}$ is taken into account, SFR models with constant efficiency per free-fall time produce too-shallow Kennicutt-Schmidt and Ostriker-Kim relations compared to observations.

\begin{figure*}
	\includegraphics[width=\linewidth]{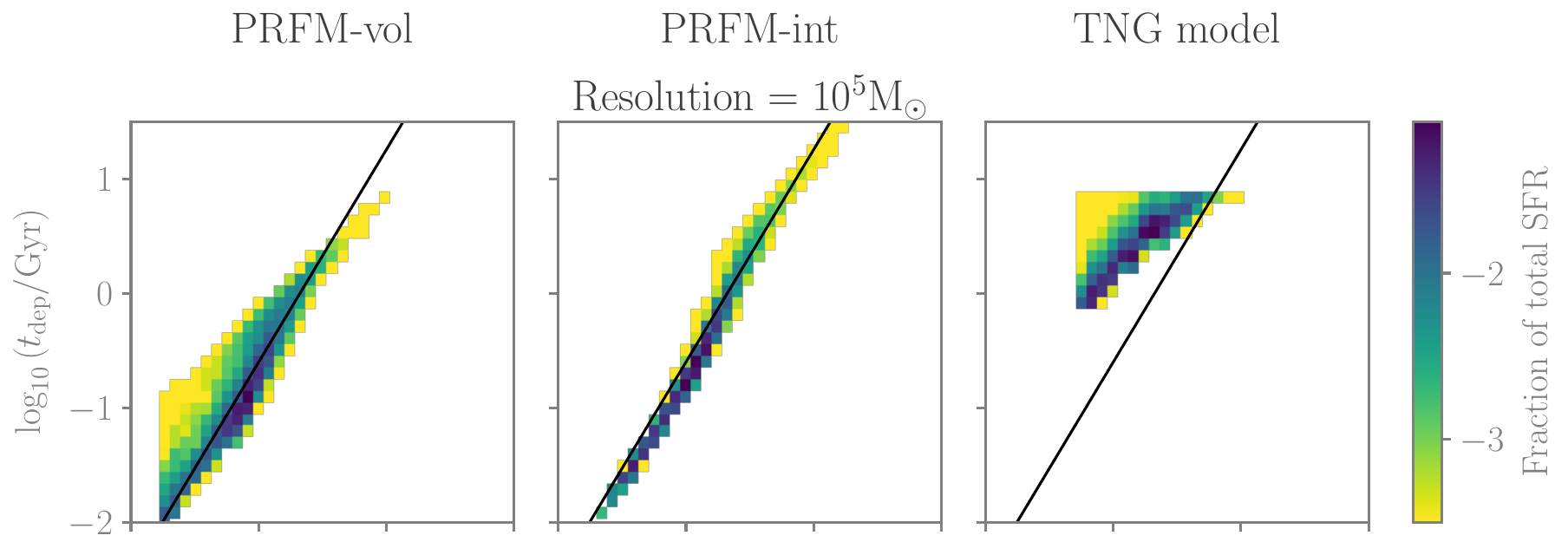}
    \includegraphics[width=\linewidth]{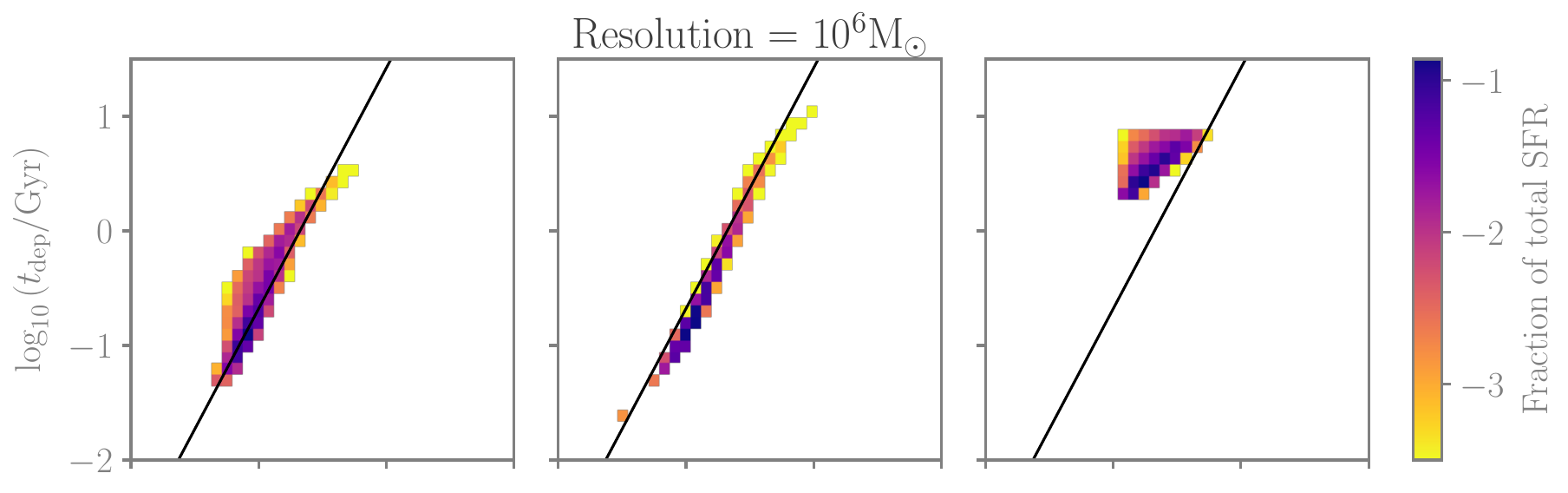}
    \includegraphics[width=\linewidth]{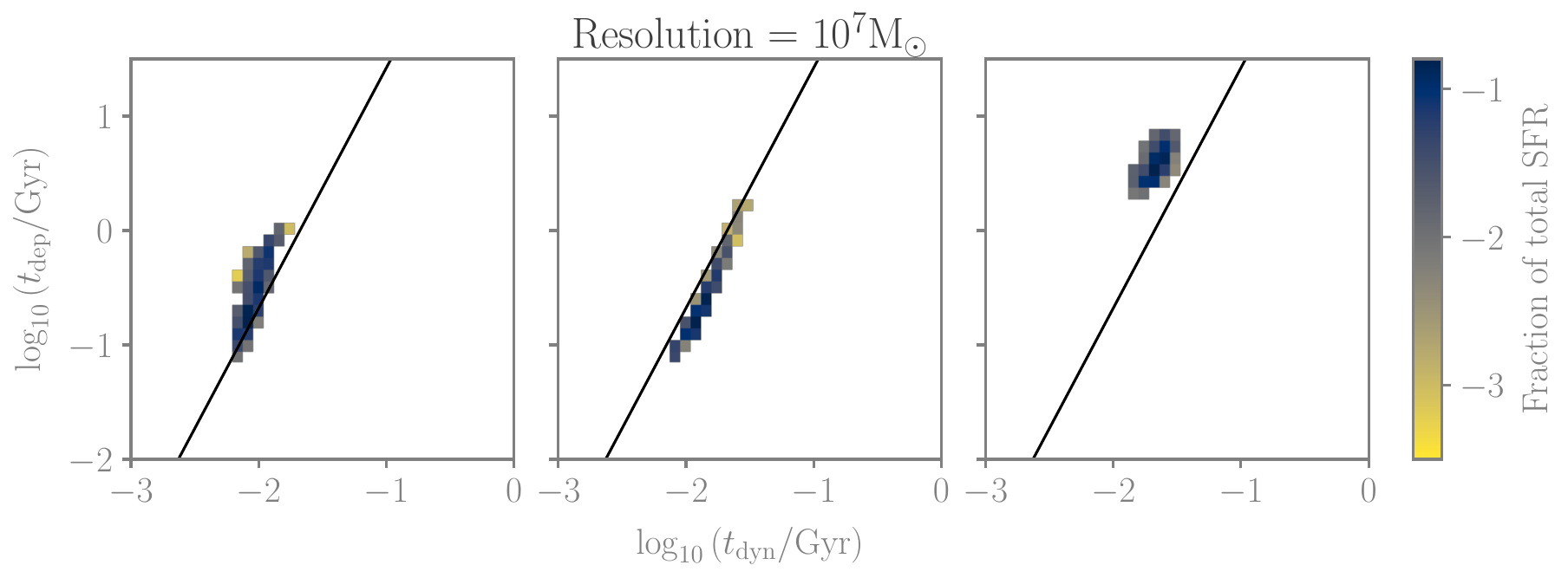}
	\caption{Distribution of gas cells in the plane spanned by the gas cell dynamical time $t_{\rm dyn}$ and the gas cell depletion time $t_{\rm dep} = M_{\rm g}/\dot{M}_*$, at a simulation time of $50$~Myr, at the three simulation resolutions of $10^5\Msun$ (top row), $10^6\Msun$ (center row) and $10^7\Msun$ (bottom row). 
    Computation of timescales is described in \autoref{fig:P_vs_tdep_tdyn}. 
    The black lines represent the least-squares best fit ($\log{t_{\rm dep}} \approx 2.1 \log{t_{\rm dyn}} + 3.5$), to the \PRFMR\ simulation, including all gas cells with ${\rm SFR} > 0$.}
    \label{Fig::SFR_tauDyn_tauDep}
\end{figure*}

Finally, in \autoref{Fig::SFR_tauDyn_tauDep} we compare the relations between 
the depletion times and dynamical times produced by each of the star formation implementations tested. Similarly to \autoref{fig:SSFR-P}, each row  compares the three different implementations at a given resolution, varying from $10^5\Msun$ in the top row to $10^7\Msun$ in the bottom row. The solid black lines in all panels represent the least-squares best fit to the $\log \tdep$ vs. $\log \tdyn$  relation for the \PRFMR\  simulation at a resolution of $10^5\Msun$.

We first consider $\tdep$ vs. $\tdyn$ for each of the star formation prescriptions at varying resolution.  Considering either the left or center column of \autoref{Fig::SFR_tauDyn_tauDep}, we can see that the relationship between $t_{\rm dyn}$ and $t_{\rm dep}$ is well-converged with respect to resolution.  This is as expected, given that $\tdep$ vs. $\tdyn$ in the PRFM prescription of \autoref{eq:SFRvol} depends only on the quantities $\sigma_{\rm eff}$ and $\Upsilon_{\rm tot}$ that are functions of pressure as calibrated by the TIGRESS simulations. The larger spread in  \PRFMR\ compared to \PRFMU\ distributions is because the \PRFMU\ model effectively only represents midplane conditions (hence producing less scatter) while \PRFMR\ allows for variations with height in the disk.  It is worth noting that the peak of the distribution from the highest-resolution \PRFMR\ simulation is slightly below this overall best-fit line, and the (narrower) distributions of all \PRFMU\ simulations match this offset. 

Interestingly, the most obvious difference between the three implementations is the dynamic range of the modeled time-scales. Comparing the PRFM (left and center columns) and TNG (right column) implementations, the range of depletion times obtained is up to three orders of magnitude larger with PRFM. The \PRFMU\  implementation (center column) also has a substantially larger dynamic range at intermediate resolution than does the \PRFMR\ implementation (left column).  Especially, higher values of $\tdep$ ($\sim 1-10$Gyr) are less well represented in the lower-resolution \PRFMR\ simulations. This is likely due to the fact that the gas cell densities and pressures,
which directly set $t_{\rm dep}$ in the \PRFMR\ implementation, become inaccurate at these low resolutions, and in particular the density in the simulation may fall below the threshold for star formation. The computation of $\tdyn$ and $\tdep$ based on column densities in \PRFMU\ is more robust, and allows the depletion time to maintain a greater spread of values down to $10^7\Msun$ resolution.

The relationship between $\tdep$ and $\tdyn$ in the \TNG\ prescription is clearly quite different from that based on PRFM theory, as calibrated with full-physics TIGRESS simulations.  Following the prescription of the original TNG simulations, in our \TNG\ simulations of isolated galaxies, the star formation efficiency per free-fall time (which depends only on density)  is a single constant value. As a result, there is much lower variation of $\tdep$ in the high-resolution \TNG\ simulation than in either PRFM simulation.  
At low resolution, there is even less variation in $\tdep$ in the \TNG\ simulation.  Characteristic values of $\tdep$ are overall lower in the PRFM simulations at all resolutions, compared to the \TNG.  As previously pointed out in \citet{Hassan2024}, this suggests that predictions of SFRs could potentially be significantly higher -- especially at early cosmic time -- if PRFM were adopted for cosmological star formation modeling, in comparison to previous results obtained using TNG.  

\subsection{Explaining model differences} \label{Sec::interp}
We now inspect some of the key properties of the \PRFMR\ (\autoref{Sec::PRFMR-props}) and \PRFMU\ simulations (\autoref{Sec::PRFMU-sims}), to provide further insight into the more favorable robustness to numerical resolution displayed by \PRFMU\ relative to \PRFMR\  (\autoref{Fig::SFR_vs_time} and \autoref{fig:SSFR-P}).  

\subsubsection{Properties of \PRFMR\ simulations}\label{Sec::PRFMR-props}
\begin{figure*}
    \includegraphics[width=\linewidth]{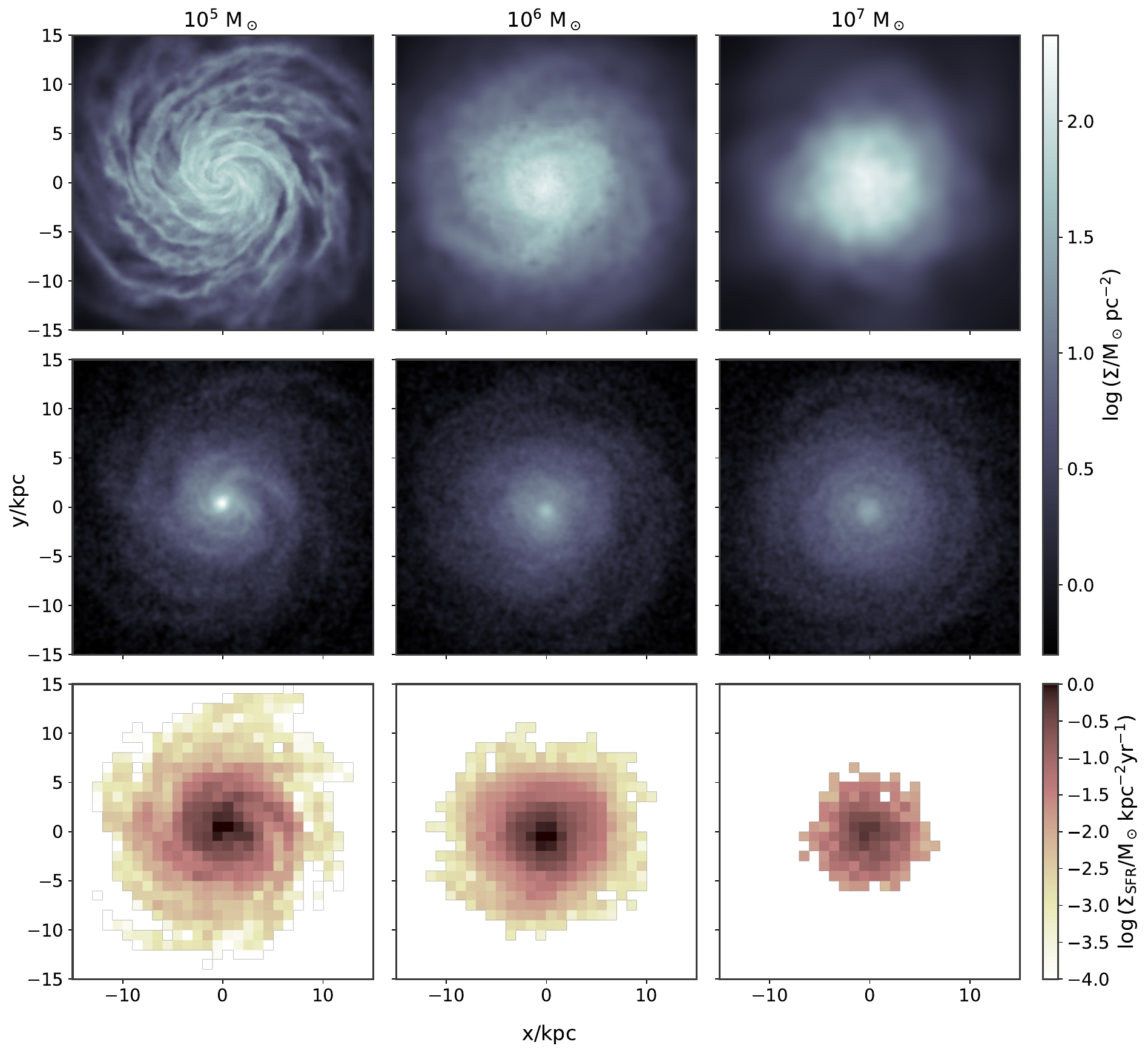}
    \caption{Maps of the gas surface density $\Sg$ (upper row), the stellar surface density $\Ss$ (center row) and the SFR surface density $\SSFR$ (lower row) for simulations adopting the \PRFMR\ implementation. Results are shown from simulations at three different mass resolutions (and corresponding typical cell size): $10^5\Msun$ ($80$~pc), $10^6\Msun$ ($190$~pc), $10^7\Msun$ ($370$~pc), at a simulation time of $50$~Myr. In the top two rows, the maps are created by summing the gas and stellar masses along columns of width $\sim 100$~pc. In the bottom row, the SFR is summed over a larger pixel size ($\sim 1$~kpc), to ensure visibility of the star-forming regions.}
    \label{Fig::PRFM-res_maps}
\end{figure*}

In \autoref{Fig::PRFM-res_maps} we show snapshots of the projected gas surface density $\Sg$ (top row), projected stellar surface density $\Ss$ (center row) and projected SFR surface density $\SSFR$ (bottom row) in the \PRFMR~simulations at resolutions of $10^5$ (left column), $10^6$ (center column) and $10^7\Msun$ (right column). The highest-resolution simulation has a flocculent spiral arm structure, visible in all three maps, which is lost as the resolution is coarsened, due to the increase in softening length from $\sim 80$~pc to $\sim 500$~pc, and finally to $\sim 1$~kpc at $10^7\Msun$ resolution. It is also apparent that the SFR is highest in the bulge region of the galaxy, due to the higher gas densities and pressures there.

\begin{figure*}
\centering
\includegraphics[width=.75\linewidth]{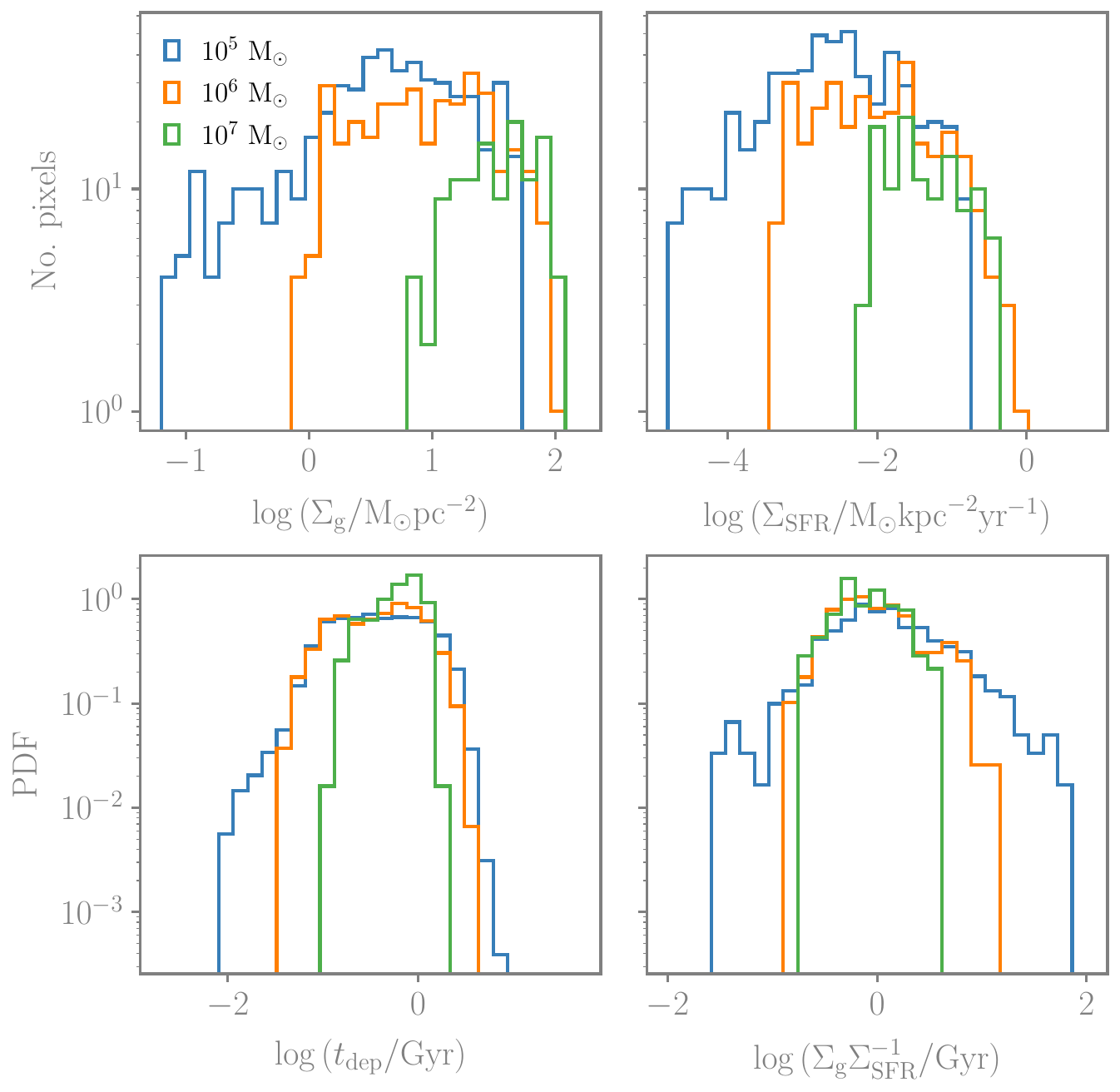}
    \caption{\textit{Upper panels:} Histograms of the gas surface density $\Sigma_{\rm g}$ and the SFR surface density $\Sigma_{\rm SFR}$, summed over columns at a scale of $1~{\rm kpc}$.  Results are from simulations adopting the \PRFMR\ implementation at resolutions of $10^5\Msun$ (blue), $10^6\Msun$ (orange) and $10^7\Msun$ (green), at a time of $50$~Myr. \textit{Lower panels:} Histograms of the depletion time $t_{\rm dep}$ computed from individual cells within the PRFM model for each simulation, and the measured ratio of $\Sigma_{\rm g}$ and $\Sigma_{\rm SFR}$ at $1~{\rm kpc}$ scale from the projected maps. Note that only star-forming gas cells contribute to the histogram of $t_{\rm dep}$, whereas all gas cells are counted in $\Sigma_{\rm g}/\Sigma_{\rm SFR}^{-1}$.}
    \label{Fig::PRFM-res-maps-hist}
\end{figure*}

In the top two panels of \autoref{Fig::PRFM-res-maps-hist}, we present 1D histograms of $\Sg$ and $\SSFR$ from the \PRFMR\ simulations.  These  quantities are presented for a resolution of $10^5$ (blue), $10^6$ (orange) and $10^7\Msun$ (green). All projections are computed over vertical columns at a horizontal scale of $1$~kpc, to make clear the differences in the distributions of the quantities at each simulation resolution.  Only gas above the density threshold $n_\mathrm{th}= 0.13 \cm^{-3}$ contributes to the $\Sg$ and $\SSFR$ histograms shown here.  

One important point to note  is that the distribution of $\Sg$ does not just narrow as the resolution is coarsened, but at $10^7\Msun$ resolution  the peak of the $\Sg$ distribution shifts upward by a factor $\sim 5$.
As a result, there is also a shift upward in the peak of the $\SSFR$ distribution by more than an order of magnitude in the $10^7\Msun$ \PRFMR\ simulation.

There are two reasons for the shift in the peak of the $\Sg$ distribution in the $10^7\Msun$ simulation; both are a result of low resolution. First, at low spatial resolution angular momentum loss in the outer part of the disk leads to a radial shrinkage, so that $\Sg$ at large radii drops and $\Sg$ at smaller radii (slightly) increases.
Second, due to poor resolution of the disk scale height (discussed further below), gas beyond $\sim 6$kpc in the $10^7\Msun$ simulation drops to a density below the star formation threshold.  
This is clear also in the maps of \autoref{Fig::PRFM-res_maps}, which show that $\SSFR$ is truncated beyond $\sim 6$kpc in the lowest-resolution simulation. As is discussed further in \autoref{app:B}, where radial profiles of surface density and midplane volume density are compared at different resolutions in \autoref{fig:radial-profiles}, the unphysically large disk thickness in the $10^7\Msun$ simulation reduces the volume density such that only the inner portion of the disk at high $\Sg$ meets the criterion $n_\mathrm{H,meas}>n_\mathrm{th}$ imposed as a star formation threshold. 

During the following discussion of our results, we will need to keep in mind the overall higher $\Sg$ for the star-forming gas at resolution of $10^7\Msun$.  It is worth noting that even in the $10^6\Msun$ simulation, star formation is truncated a point where $\Sg \sim 1 \surfunit$, above the truncation point for the $10^5\Msun$ simulation ($\Sg \sim 0.1 \surfunit$), as shown in the top-left panel of \autoref{Fig::PRFM-res-maps-hist}.  However, the $10^6\Msun$ simulation still captures  the peak of the distribution of $\SSFR$, as shown by comparison to the $10^5\Msun$ simulation in the top-right panel of \autoref{Fig::PRFM-res-maps-hist}.  
 
The lower two panels of \autoref{Fig::PRFM-res-maps-hist} compare the depletion times $t_{\rm dep}$ of individual gas cells (lower left), and $\tdep$ in projected 1~kpc regions (bottom right), where the latter is computed as the ratio of the gas to stellar surface density in post-processing. 
In both the individual-pixel and projected computations of $\tdep$, the peaks of the histograms from simulations at all resolutions are in very good agreement with each other, with the former centered at $\sim 0.5$~Gyr and the latter centered closer to $1$~Gyr since non-star-forming gas projected along the line of sight is included.  
The most noticeable feature is  a narrowing of the dynamic range at coarse resolution.  At low resolution, high density peaks cannot be resolved, and regions at low surface density have volume density that falls below the star formation threshold.  We caution, however, that while the agreement of $\tdep$ across all resolutions would naively appear to indicate convergence, this is not in fact the case.  \autoref{Fig::SFR_vs_time} already showed that global star formation histories differ in the \PRFMR\ simulations at varying resolution; we further discuss reasons for this below.  

\begin{figure*}
\includegraphics[width=\linewidth]{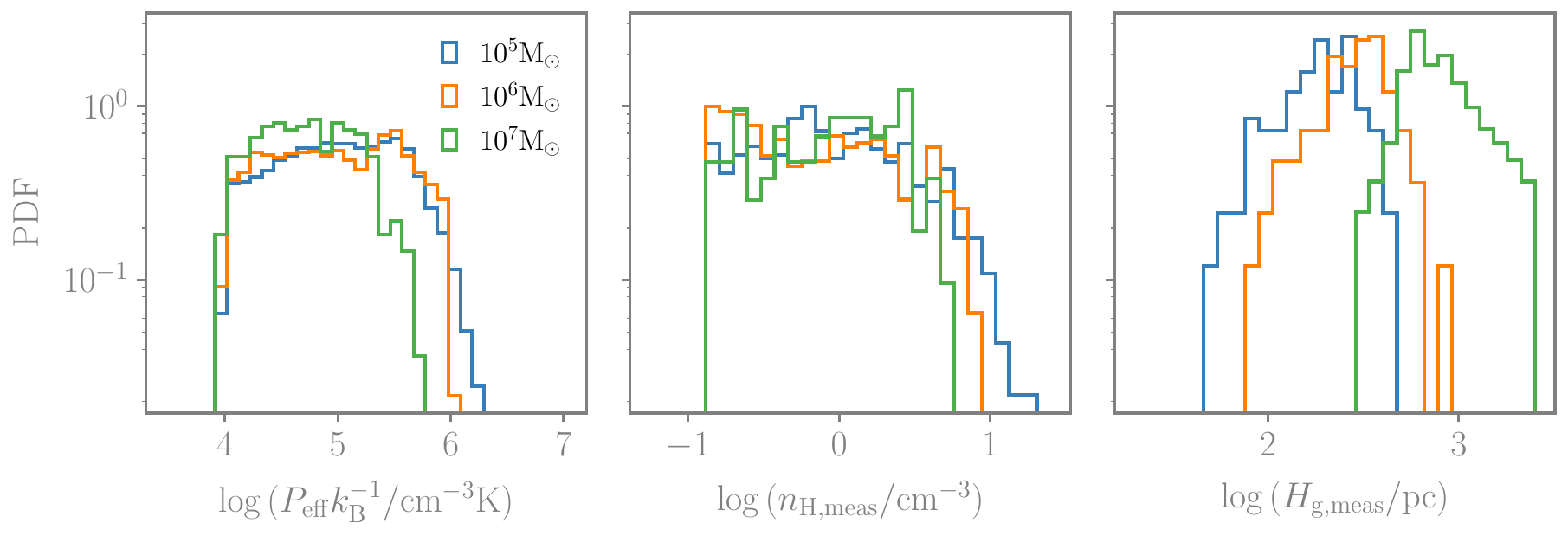}
    \caption{Histograms of the measured pressure (left), measured volume density (center), and measured scale-height (right) of star-forming gas from simulations adopting the \PRFMR\ implementation at resolutions of $10^5\Msun$ (blue), $10^6\Msun$ (orange) and $10^7\Msun$ (green). All values are those obtained directly from the simulation in post-processing.} \label{Fig::hist-res-PnH}
\end{figure*}

\begin{figure*}
    \includegraphics[width=\linewidth]{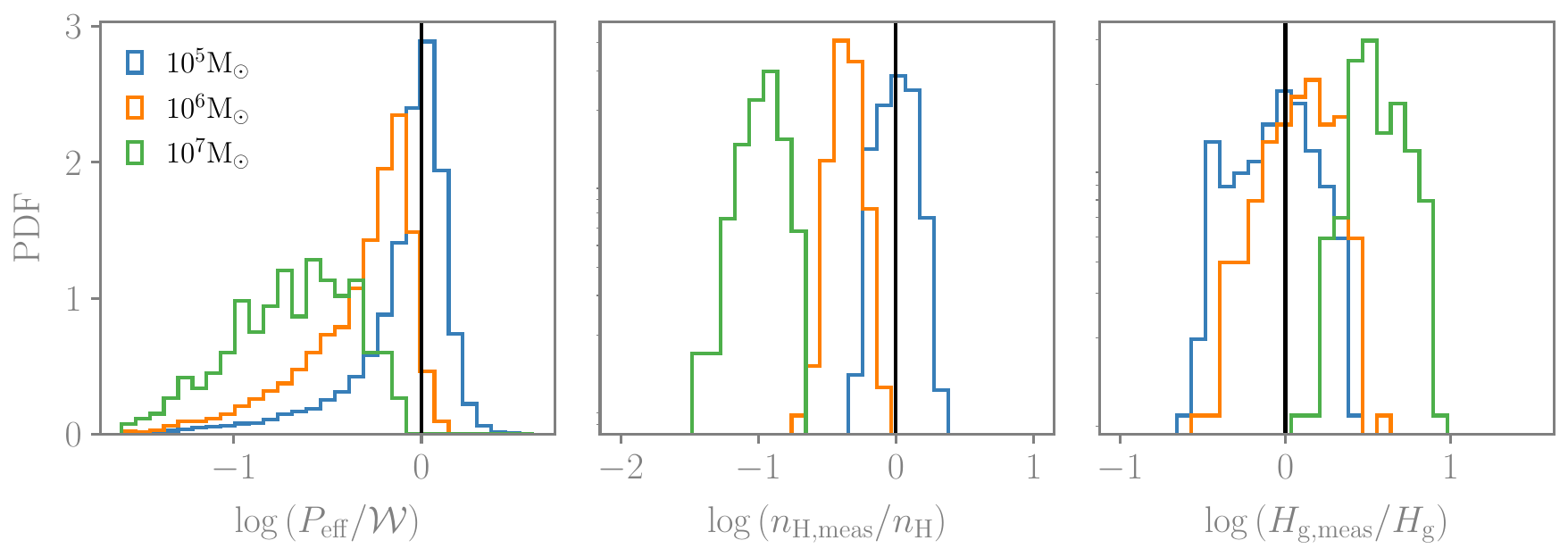}
    \caption{Comparison of measured values for ISM properties and equilibrium predictions from simulations using the \PRFMR\ implementation at a resolution of $10^5\Msun$ (blue), $10^6\Msun$ (orange) and $10^7\Msun$ (green). \textit{Left:} Ratios of the effective pressure $P_{\rm eff}$ at the midplane, computed via the EoS (\protect\autoref{Eqn::prfm-eos}) from the measured gas density, to the dynamical equilibrium pressure  computed via the ISM weight $\mathcal{W}$ (using \autoref{eq:weight_total}). \textit{Center:} Ratios of the measured mid-plane density $n_{\rm H, meas}$, to the equilibrium mid-plane density computed via the cubic equation~\protect\autoref{Eqn::Hg_cubic_resolved_Sigma} for disk thickness $\Hg$. \textit{Right:} Ratios of the measured scale-height $H_{\rm g, meas}$, to the scale-height $H_{\rm g}$ computed via the cubic ~\protect\autoref{Eqn::Hg_cubic_resolved_Sigma}. All 1~kpc averages are taken prior to computing their ratios.}
    \label{fig:P_n_H_ratios}
\end{figure*}

In \autoref{Fig::hist-res-PnH}, 
we compare for the \PRFMR\ simulations at different resolutions three of the key physical quantities that characterize the ISM and
enter either directly or indirectly in determining $\tdep$: the effective pressure $P_{\rm eff}$, the measured gas cell density $n_{\rm H, meas}$, and the measured scale-height of the gas disk, $H_{\rm g, meas}$. 
In \autoref{fig:P_n_H_ratios}, these quantities are compared to their counterparts 
that are computed based on the assumption of vertical equilibrium (as described in \autoref{Sec::resolved-case}): 
the ISM weight $\mathcal{W}$, 
the predicted mid-plane density $n_{\rm H}$, and the predicted gas disk scale-height $H_{\rm g}$. 

The most important feature to note in \autoref{Fig::hist-res-PnH} 
is the marked shift upward in the peak of the measured gas disk scale-height $H_{\rm g, meas}$ (right panel) at the lowest resolution of $10^7\Msun$ (green histograms). The shift upward in $H_{\rm g, meas}$ is direct result of the fact that it is not possible to properly resolve the ISM disk thickness in the $10^7\Msun$ simulation, given the softening length of $\sim 1~{\rm kpc}$. 
While the central panel shows that the histograms of $n_{\rm H, meas}$ are similar across all resolutions, the agreement is a signature of false rather than true convergence. The distribution of $n_{\rm H, meas}$ for the $10^7\Msun$ simulation, barring the very highest densities, agrees with high-resolution results only because the higher $\Sg$ (\autoref{Fig::PRFM-res-maps-hist}) of gas above the star formation threshold mitigates the higher $H_{\rm g, meas}$ in this poorly resolved simulation. 
As seen in the bottom row of \autoref{Fig::PRFM-res_maps}, the portion of the disk that is considered star-forming in the lowest-resolution simulation is much smaller than that in the higher-resolution counterparts, such that low surface density gas is excluded. 
If this gas at large radii had been included,
the distribution of $n_{\rm H, meas}$ would be shifted downwards at $10^7\Msun$.  In the left panel of \autoref{Fig::hist-res-PnH}, the effective pressure $P_{\rm eff}$ distribution is offset downward 
for the $10^7\Msun$ simulation compared to the 
others; this shift would be more pronounced if lower-$\Sg$ gas had been included.  
Thus, two compensating effects of poor numerical resolution in the $10^7\Msun$ simulation (a radial cutoff in star formation and a vertically puffed-up gas disk) end up producing similar  $n_\mathrm{H,meas}$ and  $\tdep$ values for star-forming gas to those in its better-resolved counterparts. 
This apparent convergence in the pressure and density distributions vanishes when the measured quantities are compared with the predicted quantities from equilibrium.

\autoref{fig:P_n_H_ratios} demonstrates the resolution dependence of the quantities $P_{\rm eff}$, $n_{\rm H, meas}$ and $H_{\rm g, meas}$ in the \PRFMR\ simulation directly, by showing values from each midplane cell relative to the computed equilibrium predictions  $\mathcal{W}$, $n_{\rm H}$ and $H_{\rm g}$. 
As in \autoref{Fig::hist-res-PnH}, only star-forming gas is included here. 
For the highest resolution, measurements and equilibrium predictions are in good agreement.  As the resolution is coarsened, however, the measured  pressures and densities decrease relative to the predicted equilibrium, while the measured gas disk scale height increases. The effect is an order of magnitude in the density, and a factor of $\sim 3-5$ for pressure and scale height, revealing 
a failure of resolution convergence of these quantities that is not directly evident in \autoref{Fig::hist-res-PnH}.

\subsubsection{Properties of \PRFMU\ simulations}
\label{Sec::PRFMU-sims}
\begin{figure*}
    \includegraphics[width=\linewidth]{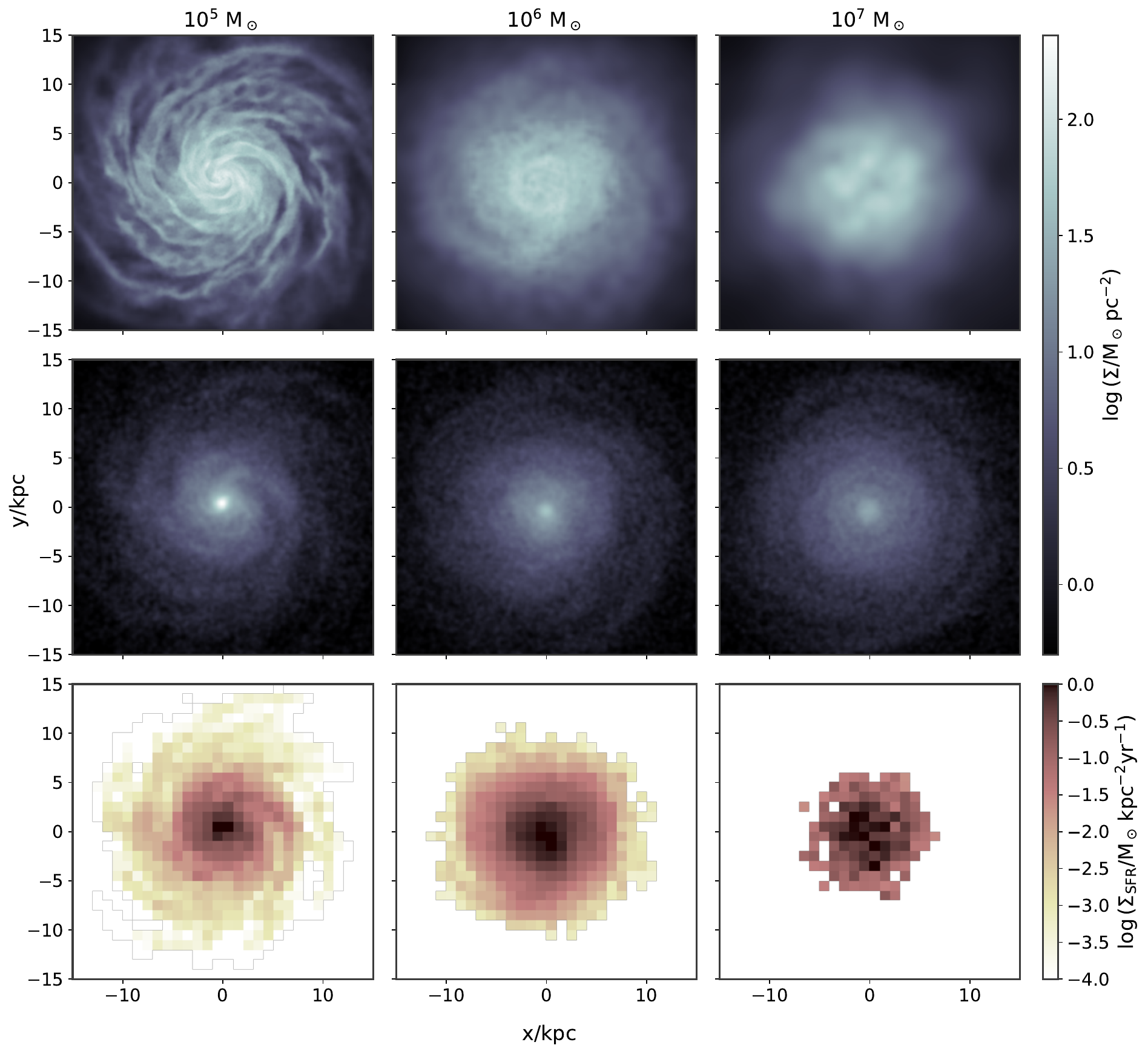}
    \caption{Maps of $\Sg$, $\Ss$, and $\SSFR$ as in \autoref{Fig::PRFM-res_maps}, but now for \PRFMU\ simulations.}
 \label{Fig::PRFM-unres_maps}
\end{figure*}

\begin{figure*}
\centering
\includegraphics[width=.75\linewidth]{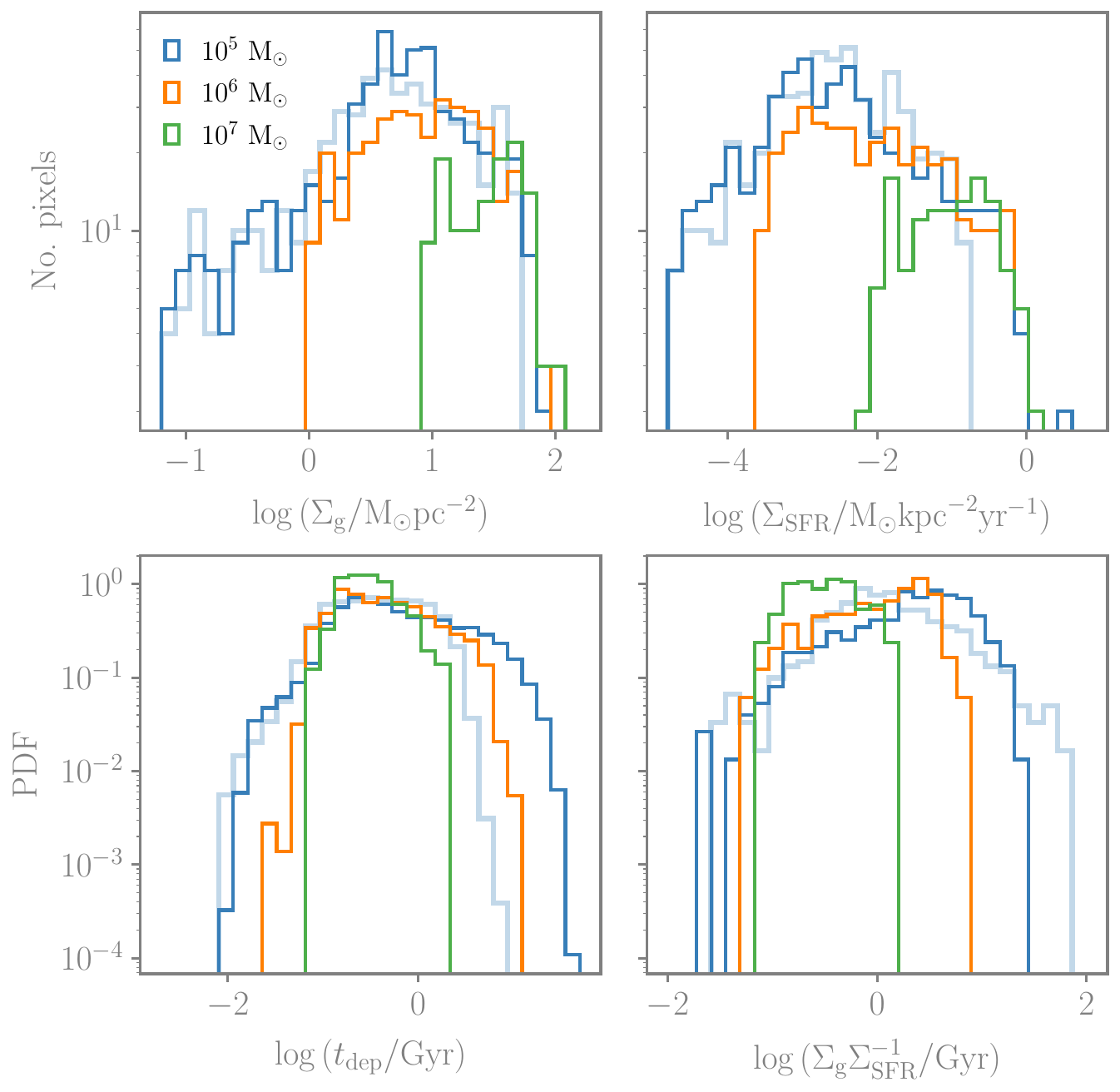}
    \caption{
    Histograms of $\Sigma_{\rm g}$, $\Sigma_{\rm SFR}$, and $\tdep$ as in \autoref{Fig::PRFM-res-maps-hist},
    but now for \PRFMU\ simulations. Transparent blue lines give the corresponding histograms from the \PRFMR\ simulations at resolution of $10^5\Msun$}
    \label{Fig::PRFM-unres-maps-hist}
\end{figure*}

\begin{figure*}
    \includegraphics[width=\linewidth]{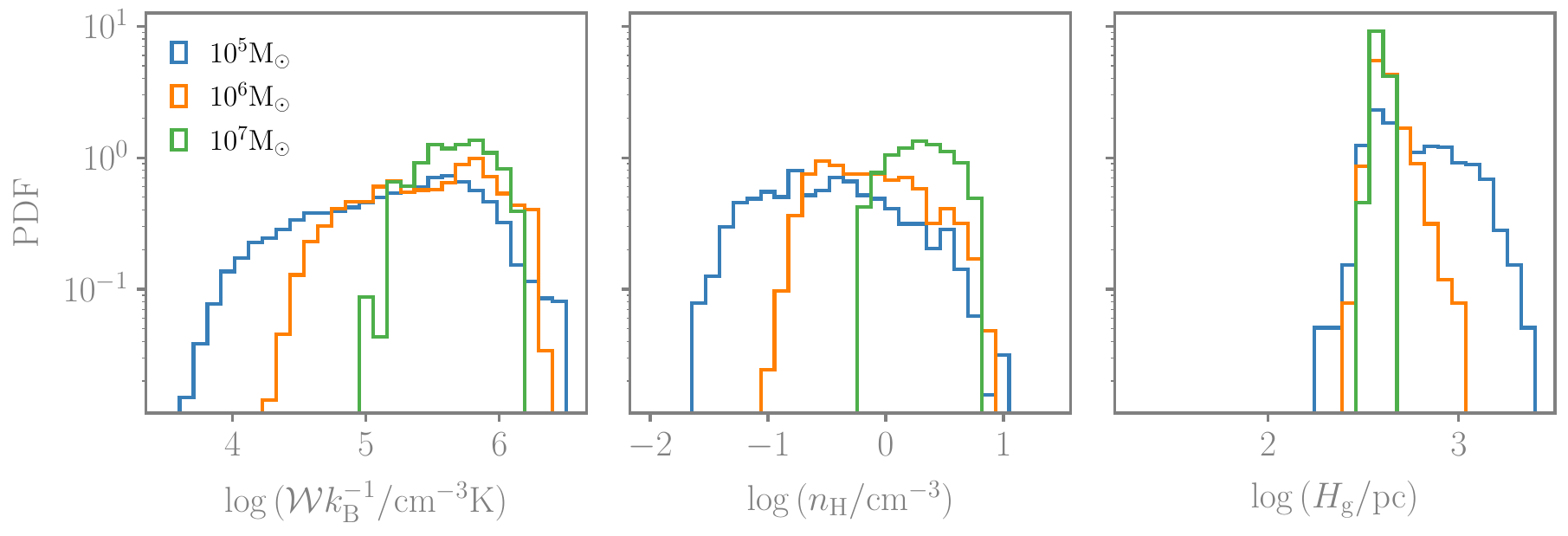}
    \caption{Histograms of the computed ISM weight $\mathcal{W}$ from \autoref{eq:P_unresolved} (left), computed equilibrium mid-plane volume density $n_H$ from \autoref{eq:n_H_equil_def} (center), and computed equilibrium scale-height $\Hg$ from \autoref{eq:Hg_unresolved} (right) for the \PRFMU\  implementation at resolutions of $10^5\Msun$ (blue), $10^6\Msun$ (orange) and $10^7\Msun$ (green).
    }
    \label{Fig:hist-unres-WnH}
\end{figure*}

We demonstrated in the previous three sections that the \PRFMR\ simulations display significantly different 
dynamical times and depletion times relative to the \TNG\ simulation, but also suffer from sensitivity to resolution. \autoref{fig:P_n_H_ratios} shows the progressive departure from expectations as resolution drops, which is result of increasing inability to vertically resolve the disk scale height. Making use of \autoref{Fig::PRFM-unres_maps}--\autoref{Fig:hist-unres-WnH}, 
we show that this problem can be alleviated using the \PRFMU\ implementation.

Comparing the projected gas surface density ($\Sigma_{\rm g}$, top row), stellar surface density ($\Sigma_*$, center row) and SFR surface density ($\Sigma_{\rm SFR}$, bottom row) maps from \PRFMU\ in \autoref{Fig::PRFM-unres_maps} to the corresponding maps in \autoref{Fig::PRFM-res_maps} for \PRFMR, we see that at resolutions of $10^5$ and $10^6\Msun$ (left and center columns), there is a quite good morphological agreement between the two implementations. The greatest difference can be seen in the simulation at $10^7\Msun$ resolution, for which \PRFMU\ displays a higher rate of star formation across its star-forming gas distribution.

The good agreement between \PRFMU\ and \PRFMR\ at $10^5\Msun$ resolution is similarly clear in the histograms of $\Sg$ (upper left panel), $\SSFR$ (upper right panel), and the depletion time $t_{\rm dep}$ (lower two panels), shown with blue transparent and solid lines, respectively, in \autoref{Fig::PRFM-unres-maps-hist}.  At  $10^6\Msun$ resolution (orange lines), 
very low values of $\Sg$  and $\SSFR$, and very high values of $\tdep$, are missing relative to the $10^5\Msun$ simulation, but otherwise the distributions are quite similar (as for \PRFMR\ in \autoref{Fig::hist-res-PnH}). 

Similar to the situation for the \PRFMR\ simulations,  \autoref{Fig::PRFM-unres_maps} shows that the star-forming disk is much smaller in the $10^7\Msun$ \PRFMU\ simulation than its higher-resolution counterparts.  This results in shifts upward in the distributions of $\Sg$ and $\SSFR$ for star-forming gas compared to the higher resolution \PRFMU\ simulations, as  shown in the top row of \autoref{Fig::PRFM-unres-maps-hist} (similar to \autoref{Fig::hist-res-PnH}). The distribution of $\SSFR$ for the $10^7\Msun$ simulation is shifted to slightly higher values in the \PRFMU\ compared to the \PRFMR\ simulation.  
Also, the distributions of $\tdep$ for the $10^7\Msun$ \PRFMU\ simulation in the two lower panels of \autoref{Fig::PRFM-unres-maps-hist} are shifted to slightly lower values than the corresponding $10^7\Msun$ \PRFMR\ simulation shown in \autoref{Fig::PRFM-res-maps-hist}.  

The cause for the downward shift in 
$\tdep$ for \PRFMU\ compared to \PRFMR\ at $10^7 \Msun$ is evident in the central panel of \autoref{Fig:hist-unres-WnH}, which shows that the computed mid-plane density $n_{\rm H}$ at $10^7\Msun$ is enhanced compared to $n_\mathrm{H,meas}$ in the $10^7$ \PRFMR\ simulation.  Although the range of $\Sg$ for star-forming gas is similar at $10^7 \Msun$ in \PRFMR\ and \PRFMU\ (see top-left panels of \autoref{Fig::PRFM-res-maps-hist} and \autoref{Fig::PRFM-unres-maps-hist}), the density computed based on a theoretical estimate of $\Hg$ in \PRFMU\ is higher than the measured density in \PRFMR\ when the disk scale height is not properly resolved.

Histograms of  the computed pressure $\mathcal{W}$ (left panel of \autoref{Fig:hist-unres-WnH}) and the gas disk scale-height $H_{\rm g}$ (right panel of \autoref{Fig:hist-unres-WnH}) at $10^7 \Msun$ are also somewhat shifted  with respect to the higher-resolution \PRFMU\ simulations.  This is entirely consistent with expectations from \autoref{eq:Hg_unresolved} and \autoref{eq:P_unresolved}, given the larger values of $\Sg$ for gas selected as star-forming in the $10^7\Msun$ simulation.

The reason that the star-forming gas in the $10^7 \Msun$ \PRFMU\ simulation has larger values of $\Sg$ than its higher resolution counterparts  (\autoref{Fig::PRFM-unres-maps-hist}) is the same as for the \PRFMR\ simulations: only gas with volume density above a certain threshold is permitted to form stars.  Since this criterion is imposed for \textit{measured} density $n_\mathrm{H,meas}$ rather than \textit{computed} density $n_\mathrm{H}$ (from \autoref{eq:n_H_equil_def}), the region of a disk that is eligible for star formation will shrink at low resolution.  This effect is shown in \autoref{fig:radial-profiles} in \autoref{app:B}, where it is clear why star formation is truncated at about $R\sim 6$kpc in both the \PRFMU\ and \PRFMR\ simulations at  $10^7 \Msun$.  In principle, it would be possible to reduce the measured density $n_\mathrm{H,meas}$ threshold for star formation to a lower level, while still only allowing star formation to occur when the computed density $n_\mathrm{H}$ exceeds a specified value.  A model refinement of this kind, 
which could be explored in the future, would allow low-level star formation to occur further out in a disk (cf. the density profile comparison for \PRFMU\ in the lower-right of \autoref{fig:radial-profiles}).  In this way, the dynamic range of $\SSFR$ would be increased, since star formation would be permitted at lower values of $\mathcal{W}$.  With a refinement of this kind, it may be possible 
for the range in $\SSFR$ in \autoref{fig:SSFR-P} for the $10^7 \Msun$  simulation with \PRFMU\ to match that of the $10^5 \Msun$ simulation.   

Even though the presently-imposed density threshold of $n_\mathrm{th} = 0.13\, {\rm cm}^{-3}$ does not permit outer disk star formation in the  $10^7 \Msun$ \PRFMU\ simulation, the global star formation history for this model is still in quite good agreement with those of $10^5 \Msun$ and $10^6 \Msun$\PRFMU\ simulations, as  \autoref{Fig::SFR_vs_time} shows.

\section{Summary and conclusions}\label{sec:summary}

In this paper, we have introduced new subgrid models for prescribing the EoS and SFR in galaxy simulations where it is not possible to directly follow the detailed physics of the multiphase, star-forming ISM.  The underpinning of these subgrid models is the PRFM star formation theory \citep{Ostriker2022}, in which a quasi-equilibrium thermal and dynamical ISM state is set by the gravity that confines the gas, and maintained by feedback. Key elements of the theory are outlined in \autoref{sec:PRFM-theory}, while calibrations of the feedback yield $\Upstot$ and EoS based on a set of TIGRESS simulations are provided in \autoref{sec:TIGRESS-calib}.   

We present two versions of the PRFM subgrid model for implementation in numerical simulations. The first version, appropriate for numerical parameter choices such that the gas disk scale height and mid-plane density are resolved  (typically mass resolution $\sim 10^5\Msun$ in a Lagrangian code), is referred to as \PRFMR, and described in \autoref{Sec::resolved-case}.  The second version, appropriate for numerical parameter choices such that the equilibrium ISM scale height would be unresolved but the radial scale length of the disk is resolved (typically  mass resolution $\gtrsim 10^6\Msun$ in a Lagrangian code), is  referred to as \PRFMU, and described in \autoref{Sec::unresolved-case}.

In \PRFMR, the density directly measured in each simulation cell goes into the EoS in order to compute the effective pressure $\Peff$.  This $\Peff$ is an input to functions that compute $\Upstot$ and the effective velocity dispersion $\seff$ (based on previous TIGRESS calibrations). The  
dynamical time $\tdyn$
in star-forming cells depends on local gas, stellar, and dark matter densities, with depletion time $\tdep =\tdyn \Upstot/\seff$.   

In \PRFMU, consideration of vertical equilibrium instead is employed to calculate the weight $\mathcal{W}$ of the ISM, based on the gas surface density $\Sg$ obtained by integrating density perpendicular to the disk.  The value of $\mathcal{W}$ is then used (instead of the measured pressure in a cell) in order to obtain equilibrium estimates of the density $n_\mathrm{H}$ and gas scale height $H_\mathrm{g}$, and from these $\tdyn$ and $\tdep$ are calculated.  

We note that in practice, either the \PRFMR\ or the  \PRFMU\ implementation can be used in any galaxy simulation, such that it is possible to  intercompare results.  Given the novelty of the technical approach in \PRFMU, it is important to confirm that at high resolution, results from a simulation applying \PRFMU\ agree with those from a simulation at the same resolution applying \PRFMR. It is also important to confirm that \PRFMU\ is robust to changes in resolution.  Additionally, it is quite interesting to test what happens when \PRFMR\ is applied at resolution that is too coarse to properly resolve the disk's scale height.  We conduct all three kinds of tests here.     

After introducing our two new models, we present results from simulations of an isolated, Milky-Way-like galaxy  at $10^5$, $10^6$ and $10^7\Msun$ resolution for both implementations, conducted using the moving-mesh code Arepo. 
For comparison with the \PRFMR\ and \PRFMU\ simulations, we also show results from simulations adopting the same initial conditions but employing the \TNG\ prescriptions for the EoS and SFR.

Our major conclusions can be summarized as follows:

\begin{itemize}

\item The \PRFMR\ and \PRFMU\ implementations at $10^5\Msun$ resolution are in good agreement with each other, in terms of the global star formation history (\autoref{Fig::SFR_vs_time}), 
galaxy morphology (\autoref{Fig::PRFM-res_maps} and 
\autoref{Fig::PRFM-unres_maps}), and distributions of gas and SFR surface densities, $\Sg$ and $\SSFR$ (\autoref{Fig::PRFM-unres-maps-hist}). Both simulations recover the targeted Ostriker-Kim relation calibrated from TIGRESS, $\SSFR \propto \Ptot^{1.2}$,  over three decades in $\SSFR$ and 
$\Ptot=\Peff$ or $\mathcal{W}$ (\autoref{fig:SSFR-P}). 

\item In comparison to TNG, the PRFM model has shorter depletion times $\tdep = M_\mathrm{g}/\dot{M}_*$, and a larger range of depletion times within a given galaxy. This is because the star formation efficiency per dynamical time $\varepsilon_\mathrm{dyn}$ in the PRFM model is not constant, but depends on the local effective pressure. Where $\Peff$ is higher (in regions of high gas, stellar, and/or dark matter density), the star formation efficiency increases. \autoref{Fig::1e5_taurat} shows the dependence of this efficiency, $\varepsilon_\mathrm{dyn}=\tdyn/\tdep= \seff/\Upstot$, on pressure.  The greater sensitivity of $\tdep$ to local conditions, under the PRFM model, could potentially alter the properties of high-redshift galaxies, compared to results obtained using traditional constant-efficiency subgrid star formation models. We refer to \citet{Hassan2024} for additional comparison to local-universe observations and potential implications of having $\varepsilon_\mathrm{dyn}$ increase in high-pressure environments.

\item At high resolution, expectations based on vertical equilibrium are satisfied in our simulations.  At low numerical resolution, however, it is not possible to capture the true density and pressure, because the gas disk scale-height is unresolved (\autoref{fig:P_n_H_ratios}). For example, at $10^7\Msun$ resolution the typical measured midplane density in the \PRFMR\ simulations is an order of magnitude lower than the equilibrium expectation (see also \autoref{fig:radial-profiles}).  This implies that any star formation model for which $\tdep$ depends only on the local volume density will not be robust to changes in resolution. Since densities are underestimated at low resolution, depletion times will be overestimated.  

\item The \PRFMU\ implementation is able to capture the targeted Ostriker-Kim power-law relation between $\SSFR$ and $\Ptot$ at all resolutions (\autoref{fig:SSFR-P}). The robustness of the \PRFMU\ subgrid model enables the \PRFMU\ simulations at resolution $10^6$ and $10^7\Msun$ to track the same global star formation history as both the \PRFMU\ and \PRFMR\ simulations at resolution $10^5\Msun$ (\autoref{Fig::SFR_vs_time}).  Both the \PRFMR\ and \TNG\ simulations, in contrast, have global star formation histories that are sensitive to numerical resolution, with the  $10^6$ and $10^7\Msun$ simulations failing to follow their  $10^5\Msun$ counterparts.
\end{itemize}

Based on the results presented here, it will be quite interesting to further test both the \PRFMR\ and \PRFMU\ subgrid model implementations in cosmological simulations of galaxy formation. The former is particularly well suited for simulations at high resolution ($\sim 10^5 \Msun$) in modest-sized comoving volumes, multi-zoom simulations in larger boxes, or for simulations that are focused on high redshift systems.  The latter is promising for enabling large suites of galaxy formation simulations at lower resolution  ($\sim 10^7 \Msun$) in larger boxes, as are needed to explore the effects on galaxy formation of varying cosmological parameters.  

We emphasize that the frameworks developed here for \PRFMR\ and \PRFMU\ are quite general, and can be implemented as subgrid models in any cosmological galaxy formation simulation code.  Here we adopt particular calibrations for $\Upstot$ and $\seff$ that are functions only of $\Ptot\approx\W$ and are appropriate to Milky-Way-like conditions, but metallicity-dependent fits calibrated from an extended set of TIGRESS simulations are also available (see \autoref{app:A}).  As the 
parameter space of  star-forming ISM numerical investigations  expands, the framework developed here provides a straightforward means to deploy calibrated results 
from future ISM studies in improving theoretical modeling of galaxy formation.

To put the present work in context, it is worth emphasizing that prescriptions for the SFR and EoS are only two among several subgrid models necessary in cosmological galaxy formation simulations where it is not possible to directly resolve important physical effects.  A related, but distinct, aspect of subgrid modeling is that required to treat galactic-scale outflows driven by feedback from star formation.  In high-resolution, full-physics simulations of the star-forming, multiphase ISM, the outflows generated by feedback are themselves multiphase, with most of the mass carried by warm gas and most of the energy carried by hot gas and cosmic rays \citep[e.g.][]{KimCG&Ostriker18,2018MNRAS.479.3042G,2019MNRAS.483.3363H,Kim_CG2020,2021MNRAS.508.2979P,2024ApJ...960..100S,2025ApJ...987..204S}.  Since the characteristic effective velocity dispersion of the ISM gas from the EoS (see \autoref{Eqn::sigma_eff}) is much lower than the escape speed from galactic potential wells, winds would not be a directly emergent phenomenon from simulations that adopt an ISM effective EoS.  However, any wind subgrid model in which outflow loading factors (for mass, momentum, energy, and metals) are parameterized as a function of local disk properties can be implemented in tandem with either \PRFMR\ or \PRFMU.  One such wind model is that presented in \citet{2020ApJ...903L..34K}, which provides a model for the joint distribution (in term of temperature and velocity) of launched material, calibrated from fits to a set of TIGRESS simulations \citep{Kim_CG2020} and parameterized by either $\SSFR$ or $\Peff$.  Launching prescriptions of this kind can be incorporated within cosmological model frameworks for treating multiphase winds, where special techniques are needed to address hot wind recoupling and subgrid momentum transfer between hot and cool phases \citep{2024MNRAS.527.1216S,2024MNRAS.535.3550S}.  While a wind model can be implemented in a galaxy formation simulation in tandem with the subgrid models we describe here, it would not be appropriate to implement other ``internal'' forms of feedback (such as momentum injection from supernovae) since their effects are already accounted for in the EoS.

Finally, we note that while the key elements of PRFM theory and numerical model calibrations 
have been validated in global isolated disk and cosmological zoom-in simulations that are entirely independent of TIGRESS \citep{2020MNRAS.498.3664G,2024ApJ...975..113J}, further testing under a wider range of conditions is needed.  Especially at high redshift, the prevalence of mergers and highly disturbed systems alters the overall geometry, and energy inputs from sources other than feedback may make a larger contribution to maintaining the total pressure.  Since the PRFM theory is based on principles of energy and momentum conservation that apply in gravitationally-bound gaseous systems very generally (with heating and cooling and turbulence processes well characterized), it is expected that an extension could be developed to allow for geometric and other modifications \citep[see discussion in][]{Ostriker2022}. It will be important to test, via direct simulations in disturbed galactic systems, the limits of current parameterizations of feedback yield and the EoS, and to develop extended calibrations that apply more broadly.

\begin{acknowledgments}
We are grateful to the referee for thoughtful reading and constructive comments on the manuscript. The work of SMRJ, ECO, and CGK was supported by grant 888968 from  the Simons Foundation to Princeton University for the Learning the Universe Collaboration. JB also acknowledges the Simons Foundation for grant 888972 to MPA in support of the Learning the Universe Collaboration.
\end{acknowledgments}

\appendix

\section{TIGRESS-NCR Calibration}\label{app:A}

The frameworks described in \autoref{Sec::resolved-case} 
for \PRFMR\
and \autoref{Sec::unresolved-case}   for \PRFMU\ can be applied quite generally as subgrid models in cosmological galaxy formation simulations. For the tests shown in the present paper, we adopt fits to TIGRESS simulations for $\Upstot$ and $\seff$ as presented in \citet{Ostriker2022}.  These fits, as summarized in \autoref{sec:TIGRESS-calib}, are for solar neighborhood metallicity.  More generally,  alternative fits to high-resolution simulations of the star-forming ISM could be substituted for those adopted here.  As we have emphasized, it is crucial that the fitting variable is a quantity that is robust to changes in resolution; this is why the variable we use is the total midplane pressure $\Ptot$, which is equal to the ISM weight $\W$ -- a quantity that can be calculated at run time in a simulation, and that is resolution-insensitive.  

\citet{Kim2024} considered a similar range of dynamical conditions to that in \citet{Ostriker2022}, but allowed for varying metallicity, also employing an updated version of the TIGRESS framework \citep[``TIGRESS-NCR'';][]{2023ApJ...946....3K} with direct ray-tracing radiation and photochemistry based on \citet{2023ApJS..264...10K}.  The fits obtained for feedback yield from the TIGRESS-NCR simulations are
\begin{equation}\label{eq:Ups_Kim24}
    \Upstot = \Upsilon_\mathrm{th} + \Upsilon_\mathrm{turb + mag}
\end{equation}
with 
\begin{equation}
     \Upsilon_\mathrm{th}= 390\kms \left(\frac{\Ptot}{10^4 k_{\rm B} {\rm cm}^{-3} {\rm K}}\right)^{-0.46}{Z'}^{-0.53}
\end{equation}
and 
\begin{equation}
  \Upsilon_\mathrm{turb + mag}  = 1170\kms \left(\frac{\Ptot}{10^4 k_{\rm B} {\rm cm}^{-3} {\rm K}}\right)^{-0.22}{Z'}^{-0.18}.
\end{equation}
for $Z'$ the metallicity relative to the solar neighborhood value.  The thermal yield has a steeper dependence on parameters than the turbulent and magnetic yield, but only becomes the dominant term at low metallicity and low pressure.  
A single combined fit for $\Upstot$ is also given in \citet{Kim2024}:
\begin{equation}\label{eq:Upstot_lowZ}
     \Upstot= 1650\kms \left(\frac{\Ptot}{10^4 k_{\rm B} {\rm cm}^{-3} {\rm K}}\right)^{-0.29}{Z'}^{-0.27}.
\end{equation}

The mass-weighted effective velocity dispersion fit for these simulations is
\begin{equation}\label{eq:sigma_eff_lowZ}
  \seff = 11.7 \kms\left(\frac{\Ptot}{10^4 k_{\rm B} {\rm cm}^{-3} {\rm K}}\right)^{0.12}Z'^{0.03}.
\end{equation}  

We note that the scaling of \autoref{eq:sigma_eff_lowZ} is somewhat shallower than in \autoref{Eqn::sigma_eff}, although the normalization is similar. This leads to a slightly less stiff EoS, $\Ptot \propto n^{1.3}$, compared to \autoref{Eqn::prfm-eos}.   

Combining \autoref{eq:Upstot_lowZ} with \autoref{eq:sigma_eff_lowZ}, the calibration for efficiency per dynamical time allowing for both varying metallicity and varying pressure is
\begin{equation}\label{eq:epsdyn_lowZ}
    \varepsilon_\mathrm{dyn} = 0.0071 \left(\frac{P_{\rm tot}}{10^4 k_{\rm B} {\rm cm}^{-3} {\rm K}}\right)^{0.41} Z'^{0.30},
\end{equation}
as given by Equation 28 of \citet{Kim2024}.  The scaling of $\varepsilon_\mathrm{dyn}$ with pressure is quite similar to that in \autoref{eq:epsdyn_calib}, with a slightly lower normalization. 
The main reason for the difference in feedback yield between the newer TIGRESS-NCR simulations and the previous ``TIGRESS-classic'' simulations is that the detailed treatment of photochemistry and heating in TIGRESS-NCR leads to a higher heating rate for a given FUV radiation field, compared to the heating rate coefficient adopted in the previous TIGRESS-classic simulations.  
Because of the slightly softer EoS, the dependence of $\tdep$ on parameters becomes $\tdep \propto n_H^{-0.5} [\rho_\mathrm{baryon} + (2/3) \rhod]^{-0.5}$, i.e. slightly weaker dependence on gas density than in \autoref{eq:tdep_explicit}.

\begin{figure*}
\includegraphics[width=\linewidth]{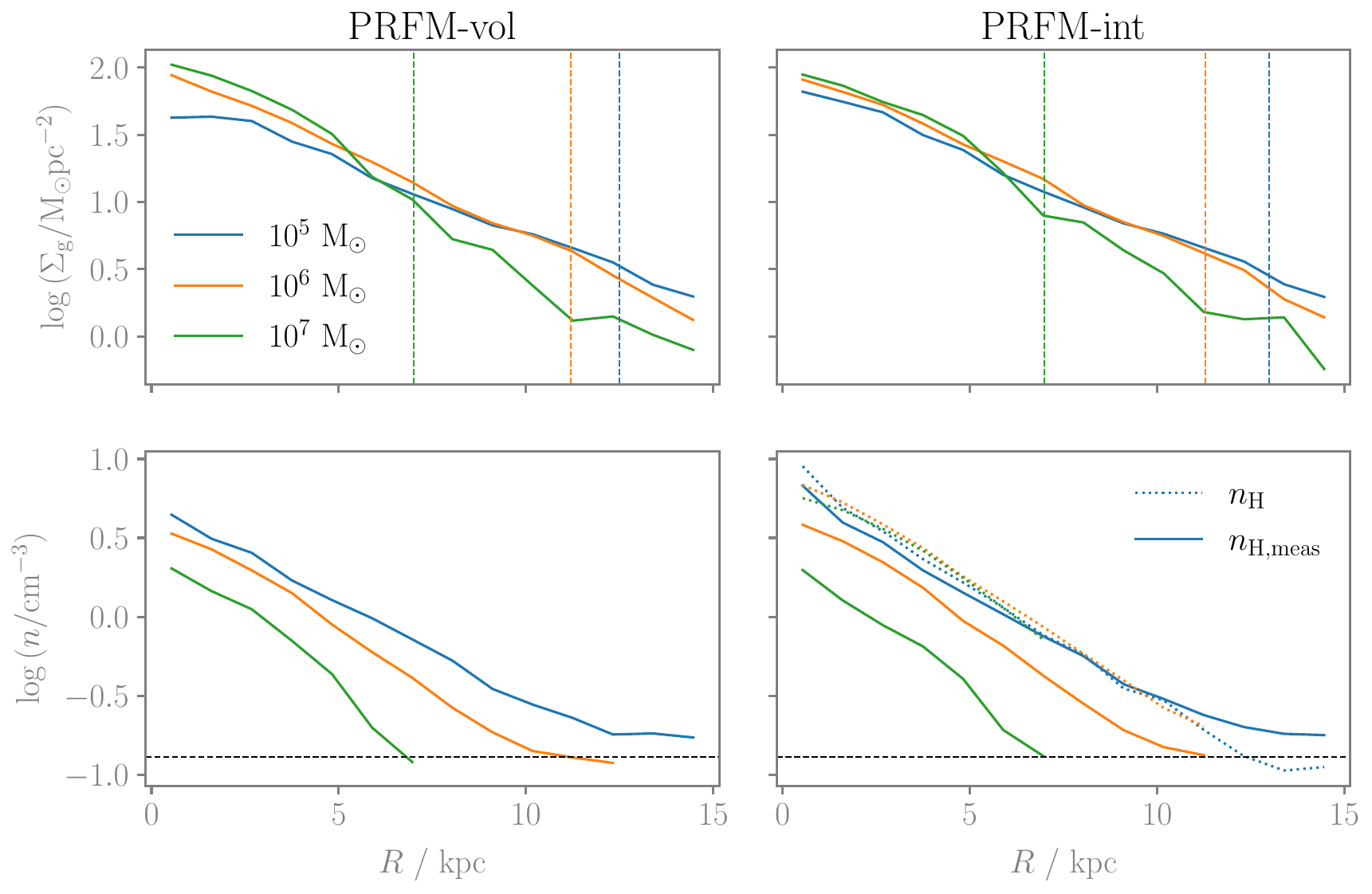}
	\caption{Radial profiles of the gas surface densities $\Sg$ (top) and volume densities $n$ (bottom) in the \PRFMR\ (left) and \PRFMU\ (right) simulations at all numerical resolutions (see key). 
    In the lower panels, solid lines show measured volume densities $n_\mathrm{H,meas}$, while the dotted lines in the lower right show calculated volume densities $n_H$.
    The dashed horizontal black lines in lower panels indicate the star formation threshold density, $n_\mathrm{th}=0.13\ \mathrm{cm}^{-3}$. The radii where this intersects with the density profiles are also marked on the $\Sg$ profiles in upper panels with vertical lines.}\label{fig:radial-profiles}
\end{figure*}

\section{Surface density and volume density profiles} \label{app:B}

As discussed in the main text, numerical resolution affects both surface density and volume density profiles in galaxy simulations.  Here, we compare profiles at the three different resolutions for the \PRFMR\ and \PRFMU\ simulations at the time, 50 Myr, which corresponds to the snapshots shown in \autoref{Fig::PRFM-res_maps} and \autoref{Fig::PRFM-unres_maps}.  These are also the snapshots used for the detailed analysis of gas and star formation properties in \autoref{Sec::PRFMR-props} and \autoref{Sec::PRFMU-sims}.  

\autoref{fig:radial-profiles} shows (with solid lines) profiles of $\Sg$ and $n_\mathrm{H,meas}$ as functions of galactocentric radius $R$ from all simulation snapshots, where $\Sg$ (top panels) is based on vertical integration and includes all gas, while  $n_\mathrm{H,meas}$ (bottom panels) is based on mass-weighted averages including only gas that is designated as star-forming. From the upper panels, it is evident that in the $10^7\Msun$ case, the gas becomes slightly more centrally-concentrated than in the higher resolution simulations, due to loss of angular momentum.  

A much more serious resolution effect is revealed in the lower panels:  for the $10^7\Msun$ simulations, the $n_\mathrm{H,meas}$ profile falls far below those of the  $10^5\Msun$ and  $10^6\Msun$ simulations, due primarily to lack of resolution of the gas scale height.  At a resolution of   $10^7\Msun$, the density chosen as a star formation threshold, $n_\mathrm{th}=0.13\ \mathrm{cm}^{-3}$, is reached at a much smaller radius ($R\sim 6$ kpc) than in the $10^5\Msun$ and  $10^6\Msun$ simulations ($R\sim 10-13$ kpc).  This is the reason for the physically smaller star-forming disks seen at low resolution in \autoref{Fig::PRFM-res_maps} and \autoref{Fig::PRFM-unres_maps}.  If the disk thickness $H_\mathrm{g,meas}$ in a simulation is (unphysically) large, a fixed $n_\mathrm{th}$ will select star-forming gas only at high $\Sg$; this situation  occurs when the equilibrium scale height $\Hg$ is unresolved in a simulation (cf. right panels of \autoref{Fig::hist-res-PnH} and \autoref{fig:P_n_H_ratios}).  The effect of only high-$\Sg$ gas satisfying the conditions for star formation at coarse resolution is evident in the upper panels of \autoref{fig:radial-profiles}, with the star-forming regime to the left of the vertical lines marked on each profile.  This is also the reason that the histograms of  $\Sg$ for star-forming gas shown in \autoref{Fig::PRFM-res-maps-hist} and \autoref{Fig::PRFM-unres-maps-hist} select only large values for the  $10^7\Msun$ simulations, significantly offset from the peak regime in the histograms of  the $10^5\Msun$ and  $10^6\Msun$ simulations. 

In the bottom-right panel, we also show with dotted lines radial profiles of the volume density $n_\mathrm{H}$ calculated based on the equilibrium scale height  (\autoref{eq:n_H_equil_def}).  The values of $n_\mathrm{H}$ at different resolution are in good agreement with each other, and are also quite comparable to the measured density $n_\mathrm{H,meas}$ in the highest-resolution simulation.   The comparison between solid and dotted green curves shows that star formation would extend to larger radius if the density calculated from vertical equilibrium considerations, $n_\mathrm{H}$, rather than density measured in the simulation, $n_\mathrm{H,meas}$, were used in the test of the condition where $n>n_\mathrm{th}$ is met. We expect that modifying the way in which gas is selected to be star-forming would therefore be able to produce larger (more realistic) stellar disks even at coarse resolution.

\bibliography{ref}
\bibliographystyle{aasjournal}

\end{document}